\documentclass[twocolumn]{emulateapj}
\usepackage{amsmath}
\usepackage{amsbsy}
\usepackage[breaklinks,colorlinks,urlcolor=blue,citecolor=blue,linkcolor=blue]{hyperref}

\newcommand{\X}{\mathrm{X}}

\begin{document}
\title{Elemental depletions in the Magellanic Clouds and the evolution of depletions with metallicity}

\author{Kirill Tchernyshyov \altaffilmark{1}, 
  {Margaret Meixner~\altaffilmark{1}$^,$\altaffilmark{2}}, 
  {Jonathan Seale~\altaffilmark{2}$^,$\altaffilmark{3}},
  {Andrew Fox~\altaffilmark{2}},
  {Scott D. Friedman~\altaffilmark{2}},
  {Eli~Dwek~\altaffilmark{4}},
  {Fr\'{e}d\'{e}ric~Galliano~\altaffilmark{5}},
  {Kenneth~Sembach~\altaffilmark{2}}}
\email{kirill@jhu.edu}
\shorttitle{Depletions in the Magellanic Clouds}
\shortauthors{Tchernyshyov et al.}

\altaffiltext{1}{Department of Physics and Astronomy, The Johns Hopkins University, 3400 North Charles Street, Baltimore, MD 21218, USA}
\altaffiltext{2}{Space Telescope Science Institute, 3700 San Martin Drive, Baltimore, MD 21218, USA}
\altaffiltext{3}{SameGrain, 3rd Floor, 101 N. Haven Street, Baltimore, MD 21224, USA}
\altaffiltext{4}{Observational Cosmology Lab, Code 665, NASA Goddard Space Flight Center, Greenbelt, MD 20771, USA}
\altaffiltext{5}{Laboratoire AIM, CEA/IRFU/Service d'Astrophysique, Universit\'{e} Paris Diderot, Bat. 709 91191 Gif-sur-Yvette, France}
\begin{abstract} 
We present a study of the composition of gas and dust in the Large and Small Magellanic Clouds (LMC and SMC, together -- the MCs) as measured by UV absorption spectroscopy.
We have measured P II and Fe II along 85 sightlines toward the MCs using archival \emph{FUSE} observations.
For 16 of those sightlines, we have measured Si II, Cr II, and Zn II from new \emph{HST} COS observations.
We have combined these measurements with H I and H$_2$ column densities and reference stellar abundances from the literature to derive gas-phase abundances, depletions, and gas-to-dust ratios (GDRs). 
80 of our 84 P measurements and 13 of our 16 Zn measurements are depleted by more than 0.1 decades, suggesting that P and Zn abundances are not accurate metallicity indicators at and above the metallicity of the SMC.
The maximum P and Zn depletions are the same in the MW, LMC, and SMC.
Si, Cr, and Fe are systematically less depleted in the SMC than in the MW or LMC.
The minimum Si depletion in the SMC is consistent with zero.
Our depletion-derived GDRs broadly agree with GDRs from the literature.
The GDR varies from location to location within a galaxy by a factor
of up to 2 in the LMC and up to 5 in the SMC. 
This variation is evidence of dust destruction and/or growth in the diffuse neutral phase of the interstellar medium.
\end{abstract}
\maketitle

\section{Introduction}
From different types of observations of dust in the Milky Way (MW) and other galaxies, we know that the gas-to-dust ratio (GDR) varies within and between galaxies. 
Dust evolution models seek to explain and quantitatively reproduce these variations. 
Most of these models treat the dust evolution problem averaged over a galaxy, and follow the basic template set by \citet{1976ARA&A..14...43A} in which different species (e.g. hydrogen or a specific metal) travels between well-mixed states (e.g. gas-phase outside of a galaxy or solid-phase in the dense molecular medium in a galaxy) according to a set of differential equations.
The current generation of dust evolution models \citep[e.g.,][]{2011ApJ...727...63D,2014A&A...562A..76Z,2014arXiv1412.2755F} include a more comprehensive and relatively well-constrained set of dust production and destruction mechanisms than the original, but are still focused on reproducing galaxy-averaged properties.
As the number of destruction and production parameters that need tuning is quite large, there is more than one way to correctly reproduce a galaxy-averaged dust-to-gas ratio.
The current-generation models listed above can all reproduce trends that have been observed between the metallicity and GDR of a galaxy while disagreeing on what the most important mechanism for dust evolution is.
One possible way to break this degeneracy is to increase the number of constraining observations by attempting to reproduce GDRs averaged over different phases of the interstellar medium (ISM) of each galaxy.

Computing an ISM-phase-averaged GDRs over a galaxy requires taking a comprehensive inventory of gas and dust over multiple phases, which requires observations of different gas mass tracers with sufficiently high spatial resolution to resolve the dense molecular phase.
Some of the most complete necessary datasets are available for the Large and Small Magellanic Clouds (LMC and SMC, together -- the MCs), which are two nearby, sub-solar metallicity ($1/2$ and $1/5$ of the solar metallicity, respectively) dwarf galaxies. 
The dust content of the MCs has been extensively studied as part of the \emph{SAGE} \citep{2006AJ....132.2268M,2011AJ....142..102G} and \emph{HERITAGE} \citep{2013AJ....146...62M} programs.
As part of these programs, dust mass \citep{2014ApJ...797...85G} and GDR \citep{2014ApJ...797...86R} maps of the LMC and SMC have been made. 

When these GDR maps are binned by ISM phase, one finds different GDRs in diffuse neutral gas and dense molecular gas. 
However, the difference between these GDR values is not quantitatively accurate.
Due to degeneracies that are explored in detail in \citet{2014ApJ...797...86R}, it is not currently possible to accurately compute dust masses in dense molecular gas from dust emission. 
One way to resolve some of these degeneracies is to measure GDRs in various environments in the LMC and SMC using a method whose systematic uncertainties are different from those of the dust emission method.
One of the aims of this study is to measure GDRs in the LMC and SMC using elemental depletions derived from ultraviolet (UV) absorption spectroscopy. 
A depletion is the difference between an element's gas-phase abundance and its intrinsic (i.e. combined gas-phase and solid-phase) abundance. 
If one makes the assumption that the missing amount is entirely in dust, which is a reasonable assumption in the molecule-poor diffuse neutral medium (DNM) through which UV spectroscopy is possible, depletions can be converted to solid-phase abundances.
Combining the solid-phase abundances of the main constituents of dust yields a GDR.

Almost all depletion studies to date have focused on the DNM of the MW.
The two main astrophysical results of the ensemble of MW depletion studies are that in a single location, elements with higher condensation temperatures tend to be more depleted \citep{1974ApJ...187..453F} and that the depletions of every element track each other and the volume density of the gas they are associated with in a consistent and continuous way \citep{Jenkins:2009ke}.
The latter result also implies that the gas volume density and GDR in a small amount of DNM are correlated.
The SMC is the next-best studied galaxy after the MW, with four sightlines along which some or all of Mg, Si, and Fe have been measured \citep{1997ApJ...489..672W,2001ApJ...554L..75W,2006ApJ...636..753S}.
These four measurements hint at possible differences between the composition of dust in the SMC and MW, but are inconclusive. 

No other system has depletion measurements of multiple important dust constituents along more than two sightlines. 
There have been several one-to-two sightline studies of local galaxies (\citealt{2014ApJ...795..109J} and references therein) and many single sightline studies of damped Lyman-$\alpha$ systems (DLAs) (\citealt{2012ApJ...755...89R} and references therein). 
Because intrinsic elemental abundances in most local galaxies and all DLAs are not available, these studies have relied on MW depletion patterns to interpret their observed gas-phase abundances. 
The present study will provide depletion patterns for two more galaxies at sub-solar metallicities, which may be more appropriate contexts for the interpretation of gas-phase abundances in low-metallicity systems.

In this paper, we present new measurements of silicon (Si II), phosphorus (P II), chromium (Cr II), iron (Fe II), and zinc (Zn II) column densities in the LMC and SMC. These measurements are made using 16 new and 84 archival spectra. Our sample, observations, and data reduction are described in Section \ref{sec:obs}. 
Our data analysis, which involves measuring column densities and computing gas-phase abundances, depletions, solid-phase abundances, and GDRs, is described in Sections \ref{sec:qa:gas} and \ref{sec:qa:dust}.
A discussion of our results and a comparison of gas-phase abundances in the MCs and DLAs is presented in Section \ref{sec:discussion}.
Our results are briefly summarized in Section \ref{sec:summary}.

\section{Observations and archival data}
\label{sec:obs}

\begin{deluxetable*}{lcccccccc}
  \tablecaption{Locations, stellar parameters, and ISM parameters of the
    COS sample}
    \tablehead{
    \colhead{Target} & \colhead{Galaxy} & \colhead{RA (J2000)} & \colhead{Dec (J2000)} & \colhead{Spectral type}  & \colhead{V}  & \colhead{E(B-V)} & \colhead{$\log(N(\text{H I}))$} & \colhead{$\log(N( \mathrm{H}_2)) $} }
    \tabletypesize{\scriptsize}
    \tablewidth{0pt}
    \startdata
  SK 9    & SMC & 11.6360 & -73.1016  & O5 V
  & 13.55 & 0.15  & 21.76 & 17.03 \\
AzV 47    & SMC & 12.2145 & -73.4329  & O8 III
& 13.44 & 0.13  & 21.32 & 18.54 \\
AzV 95    & SMC & 12.8400 & -72.7375  & O7 III
& 13.78 & 0.14  & 21.49 & 19.40 \\
AzV 238   & SMC & 14.9813 & -72.2272  & O9.5
III & 13.64 & 0.13  & 21.41 & 15.95
\\
AzV 327   & SMC & 15.7939 & -72.0373  & O9.5
II-Ibw  & 13.03 & 0.13  & 20.93 & 14.79
\\
SK 116    & SMC & 16.2323 & -72.7800  & O9
Iabw      & 12.59 & 0.13  & 21.57
& 18.53 \\
SK 143    & SMC & 17.7324 & -72.7156  & O9.7
Ib    & 12.83 & 0.36  & 21.00 &
20.93 \\
AzV 476     & SMC & 18.4269 & -73.2915  & O6.5 V
& 13.52 & 0.23  & 21.85 & 20.95\\

SK -65 22   & LMC & 75.3462 & -65.8759  & O6
Iaf+    & 12.08 & 0.11  & 20.58 &
14.93 \\
 SK -67 5     & LMC & 72.5789 & -67.6606  &
 O9.7 Ib    & 11.34 & 0.14  & 21.00
 &  19.46 \\
 SK -71 45  & LMC & 82.8154 & -71.0695  & O4-5
 III(f) & 11.54 & 0.16  & 21.09 & 18.63
 \\
SK -66 172  & LMC & 84.2725 & -66.3598  & O2 III
& 13.13 & 0.18  & 21.25 & 18.21 \\
SK -70 115  & LMC & 87.2069 & -70.0661  & O6.5
Iaf & 12.24 & 0.20  & 21.30 &
19.94\\
BI 173    & LMC & 81.7914 & -69.1324  & O8.5
II(f) & 12.96 & 0.15  & 21.34 & 15.64
\\
 BI 42  & LMC & 74.2536 & -66.4070    & O8 V
& 12.95 & 0.19  & 21.51 & 18.69 \\

SK -68 135  & LMC & 84.4547 & -68.9171  & ON9.7
Ia+ & 11.36 & 0.26  & 21.60 & 19.87
\enddata
\tablecomments{All values from \citet{Welty:2012ct} and references therein.}
\end{deluxetable*}
\label{tab:observations}

\begin{figure*}[t]
\centering
\includegraphics[width=\textwidth]{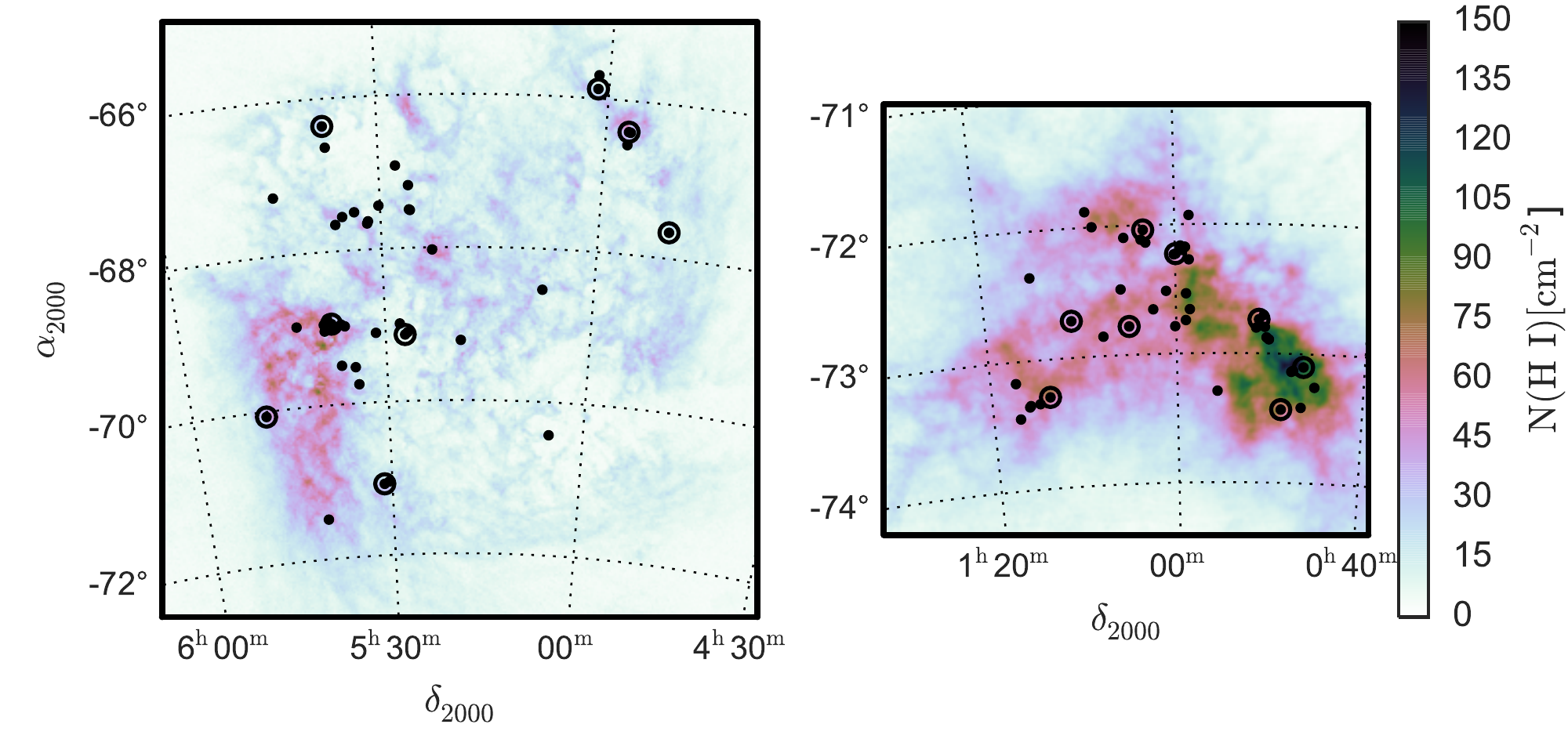}
\caption{HI column density maps of the LMC (left) and SMC (right). The positions of the 42 LMC and 46 SMC archival FUSE sightlines used in this study are marked with dots. The 8 LMC and 8 SMC sightlines that also have new COS observations are emphasized with large, open circles.}
\label{fig:sightlines}
\end{figure*}

The spectra presented in this study are a combination of new
observations from the Cosmic Origins Spectrograph
\citep[COS;][]{2012ApJ...744...60G} on the \textit{Hubble Space
  Telescope} (\textit{HST}) along 16 sight lines and supplementary archival
spectra for these 16 and 69 additional sight lines from the Far Ultraviolet
Spectroscopic Explorer
\citep[\textit{FUSE}][]{2000ApJ...538L...7S}. Figure
\ref{fig:sightlines} shows the positions of these sight lines in the
galaxies.

\subsection{COS Observations}
\label{sec:cosobs}

The COS spectra of the Magellanic Cloud sight lines presented here
were obtained between the dates of 2012 October 12 and 2013 August 19
as part of \textit{HST} program number 13004. 
All successfully acquired targets were faint
enough to use the primary science aperture, and were dithered with
all 4 FP-POS positions to improve the limiting
signal-to-noise ratio (SNR). 
During the observation of target AzV 388, HST was unable to acquire the guide star, resulting in a mis-pointing and a failed observation.
Observations were conducted with the near-UV (NUV)
G185M grating at central wavelengths of 1921 and $1953\text{ \AA}$\thickspace and a
spectral resolution of 17 km sec$^{-1}$.

\subsection{COS Targets}
\label{sec:costargets}

Table \ref{tab:observations} lists the 8 SMC and 8 LMC stars along the lines of sight
targeted in our new COS observations with their sky coordinates,
spectral types, V, E(B-V), and HI and H$_{2}$ column densities as
reported in \citet{Welty:2012ct}. These targets will be referred to as the
COS sample.

The COS sight lines were chosen from a set of 285 Magellanic Cloud targets with HI and/or H$_{2}$ column densities measured from archival \textit{HST} and \textit{FUSE} spectra presented in \citet{Welty:2012ct}. In order to explore the dependence of depletion on the column density ($N$) and phase of gas, sight lines were chosen to most completely cover the $2N$(H$_{2}$)/($N$(HI)+$2N$(H$_{2}$)) vs. $N$(HI)+$2N$(H$_{2}$) parameter space. The targets sample a range of total neutral hydrogen column densities from $N\rm{(HI)}+2N\rm{(H_{2})}$ of $\sim10^{20.5}$ to $\sim10^{22}$ cm$^{-2}$ and molecular hydrogen fraction, $2N$(H$_{2}$)/($N$(HI)+$2N$(H$_{2}$)), of $\sim 50 \%$ to $\ll 1\%$. Few sight lines with high molecular fractions exist in the original sample of 285, so our COS sample is dominated (14/16) by those with molecular fractions of $<10\%$.

\subsection{FUSE Targets}
\label{sec:fusetargets}
Data for the 69 supplementary sightlines, of which 32 are towards the LMC and
37 are towards the SMC, were downloaded from the FUSE MC Legacy Project archive \citep{2009PASP..121..634B} and analyzed with no further
processing. The FUSE MC Legacy Project archive contains spectra
towards a total of 287 stars. From these, we selected stars whose continuum SNRs
in the region from 1140 to 1155 \AA\ were greater than 5 and 
towards which \citet{Welty:2012ct} had detected H I and H$_2$. These
targets will be referred to as the FUSE-only sample.

\subsubsection{Ancillary literature column densities}
Two of the LMC and four of the SMC targets in the FUSE-only sample
have been observed and analyzed before \citep{1997ApJ...474L..95R, 
2001ApJ...554L..75W, 2006ApJ...636..753S}.
All of these targets have Zn
and Cr measurements and three of the four SMC targets have Si
measurements. 
Spectra towards Sk 155 and Sk 108 has been separately analyzed by both
\citet{2001ApJ...554L..75W} and \citet{2006ApJ...636..753S}. 
We adopt the values of the latter when they are given and the former
otherwise.
We note that this work is the first to use these
measurements to compute depletions relative to hydrogen; the original
authors did not have access to the hydrogen absorption measurements of \citet{Welty:2012ct}.

\subsection{Data Reduction}
\begin{figure*}
  \centering
  \includegraphics[width=\linewidth]{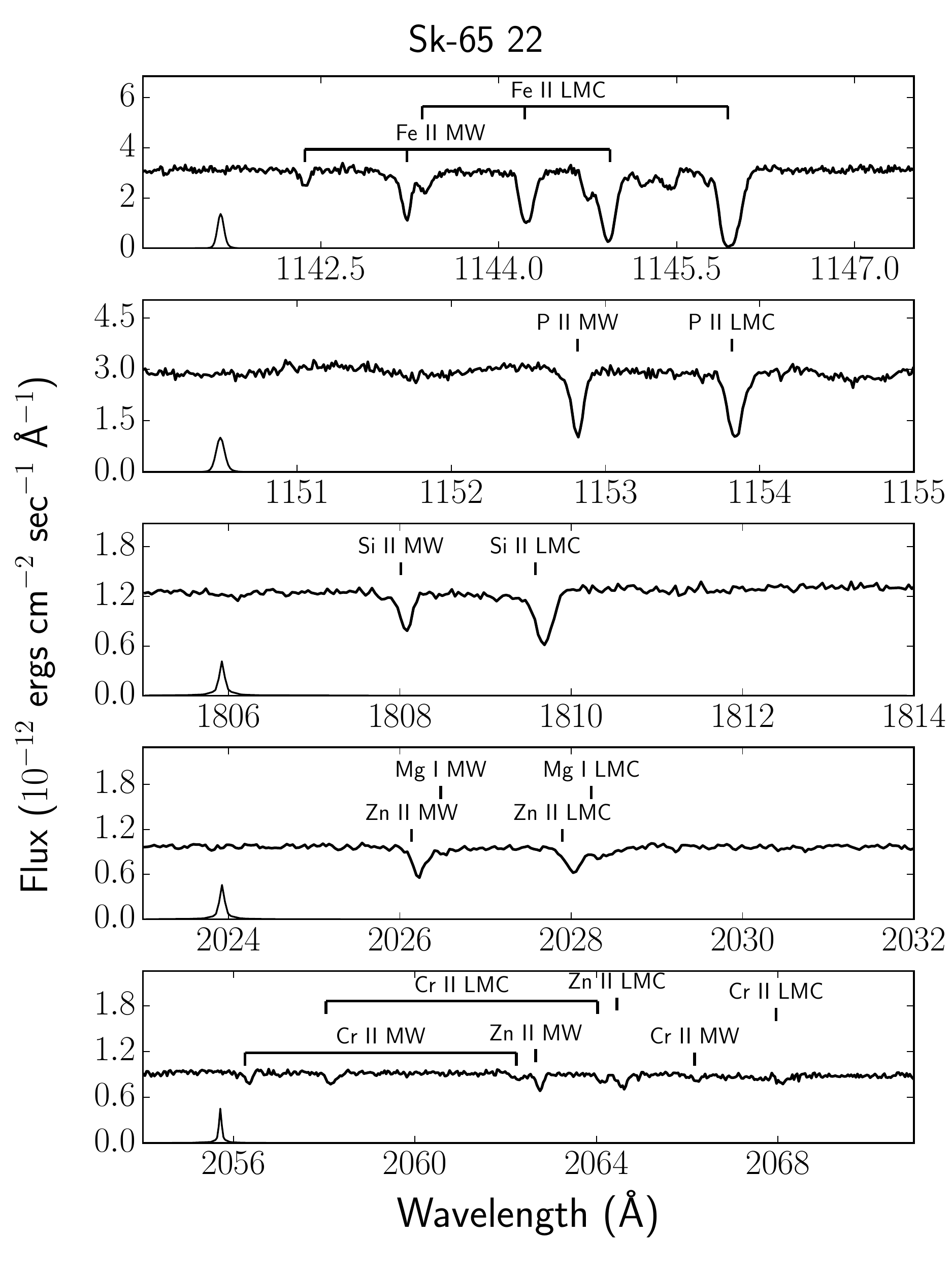}
  \caption{FUSE (top two panels) and COS (bottom three panels) observations of the LMC O6 Iaf+ star Sk -65
    22. Labels show the locations of ISM absorption
    features of interest. The instrumental LSF for each wavelength
    range is shown at the
    bottom left of each panel.}
  \label{fig:datavis:sk6522}
\end{figure*}
\begin{figure*}
  \centering
  \includegraphics[width=\linewidth]{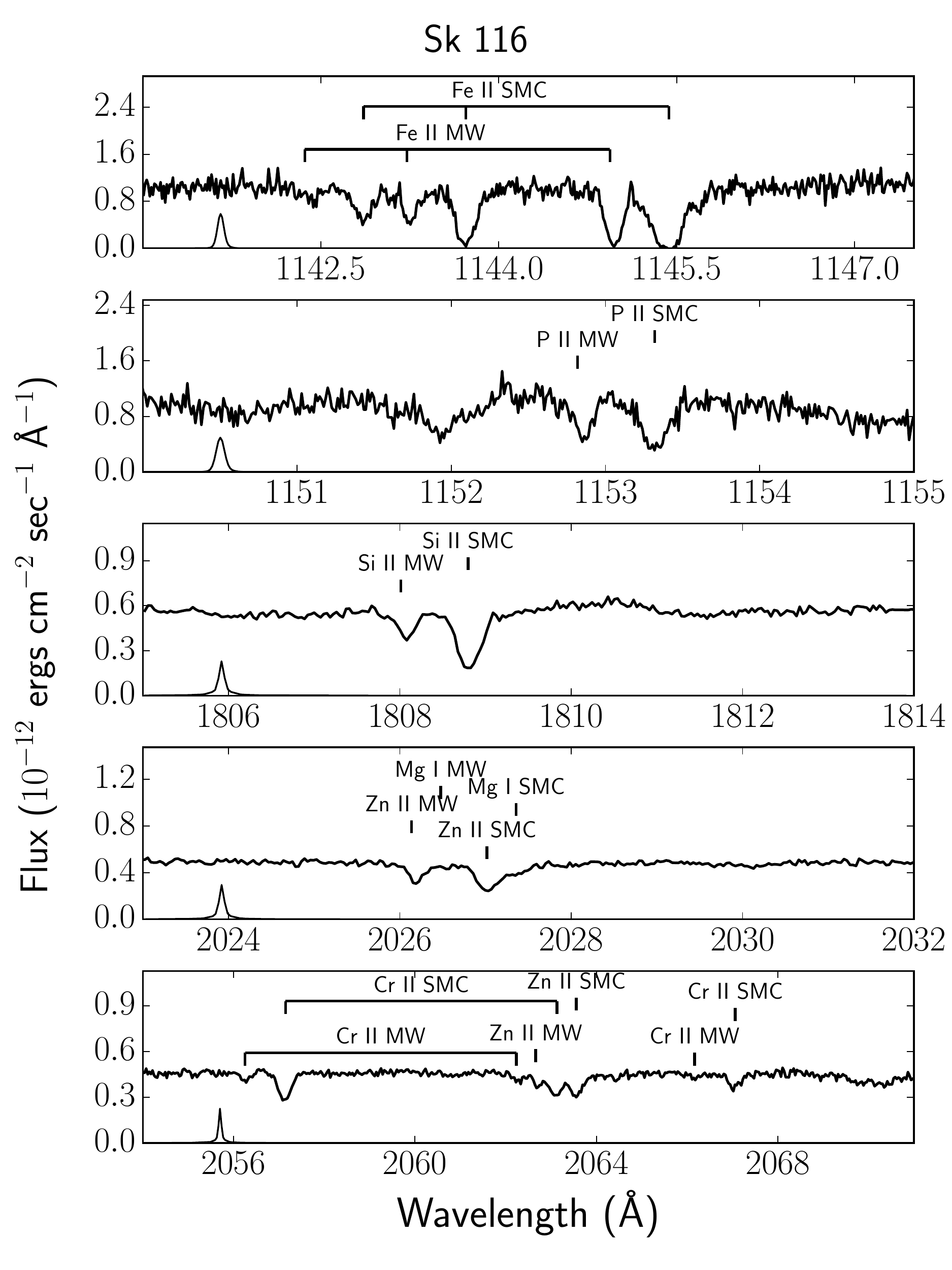}
  \caption{FUSE (top two panels) and COS (bottom three panels) observations of the SMC O9 Iabw star Sk 116. Labels
    show the locations of ISM absorption features of interest. The instrumental
    LSF for each wavelength range is shown at the bottom left of each panel.}
  \label{fig:datavis:sk116}
\end{figure*}

Figures \ref{fig:datavis:sk6522} and \ref{fig:datavis:sk116} show
spectra towards SK $-65$ 22 in the LMC and SK 116 in the SMC. Labels and vertical lines
indicate the typical wavelengths of absorption lines of interest;
FUSE and COS line spread functions (LSFs) are shown at the bottom left
of each panel \citep{2002ApJS..140...19K, Ghavamian:2009ts}. The relative velocities assumed for LMC and SMC
absorption are based on the ranges given in \cite{Brns:2005dl}. 
Most of the other spectra in
this study resemble the ones described above.

The continua of SK $-68$ 135 and SK $-65$ 22
contain broad absorption and emission lines, including some which
overlap with absorption lines of interest. In some of the FUSE
spectra, the absolute flux level
between $1140$ and $1150\text{ \AA}$\ is decreased relative to the flux level
at nearby wavelengths due to a shadow from a repeller grid falling on
part of the FUSE detector. All of our Fe II lines fall within this
region; spectra where the shadow is especially strong are only
marginally usable and have been excluded. There are
absorption lines due to gas between the MW and LMC in half of the
observations towards the LMC. This gas has relative velocities which
range from 100
to 180 $\text{ km sec}^{-1}$, has been observed before
\citep{Lehner:2009cg}, and can usually be distinguished from MW and LMC
absorption.

The resolutions of FUSE and COS are $\sim 13 \text{ km sec}^{-1}$ \citep{2002ApJS..140...19K} and
$\sim 15 \text{ km sec}^{-1}$ \citep{Ghavamian:2009ts},
respectively. Isolated lines in higher resolution
($\sim 3 \text{ km sec}^{-1}$) spectra of the diffuse neutral medium
in the MCs have widths of
order a $\text{ km sec}^{-1}$ \citep{1997ApJ...474L..95R,
  1997ApJ...489..672W}, implying that all of the absorption we see is almost certainly
unresolved. 
\section{Quantitative analysis: gas-phase}
\label{sec:qa:gas}
We compute four measurements of gas-phase metal
content, which are listed below in order of increasing distance from
the data.
From the spectra, we measure column densities for P II, Zn II, Si II,
Cr II and Fe II. 
Normalizing these column densities by the hydrogen column density gives
us ion abundances.
If we had measurements of a single element in more than one ionization
state, we would compute an ionization correction for each
sightline. Instead, we check that the ionization corrections for most
of our sightlines should  be negligible and adopt the ion abundances
and elemental abundances.
Normalizing these elemental abundances by reference, or total gas- and
solid-phase, abundances gives us elemental depletions.
Finally, we analyze the distribution of depletions in each galaxy
using the formalism from \citet{Jenkins:2009ke}.
This formalism gives us a concise summary of an element's depletion
trends each galaxy and allows us to impute partially missing data, such
as Zn, Si, and Cr depletions for sightlines in the FUSE-only sample.
Below, we describe each of these steps in moderate detail;
the detailed mechanics of steps with relatively complex implementations are
described in the appendices. 

\begin{figure*}
  \centering
  \includegraphics[width=\linewidth]{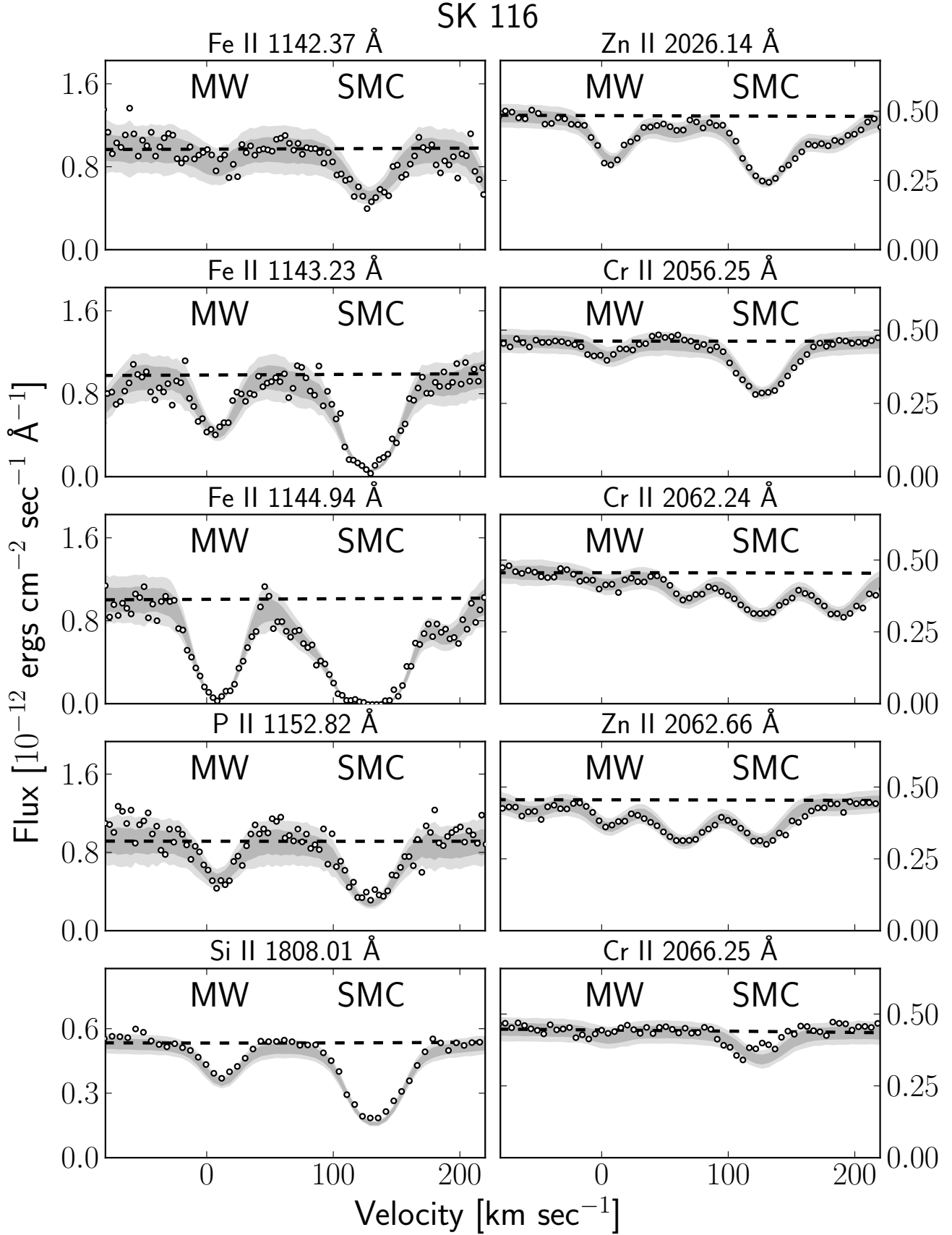}
  \caption{Profile fit of the observation towards Sk 116 in the
    SMC for the different ions. Listed next to the ions is the rest frame ($v=0$) 
    wavelength of the absorption line.  Circles represent the data; 
    light and dark grey regions are
    the 68 and 95$\%$ credible regions about the median fit; the
    dashed line is the median continuum level}
  \label{fig:fit:sk116}
\end{figure*}

\begin{figure}
  \centering
  \includegraphics[width=\linewidth]{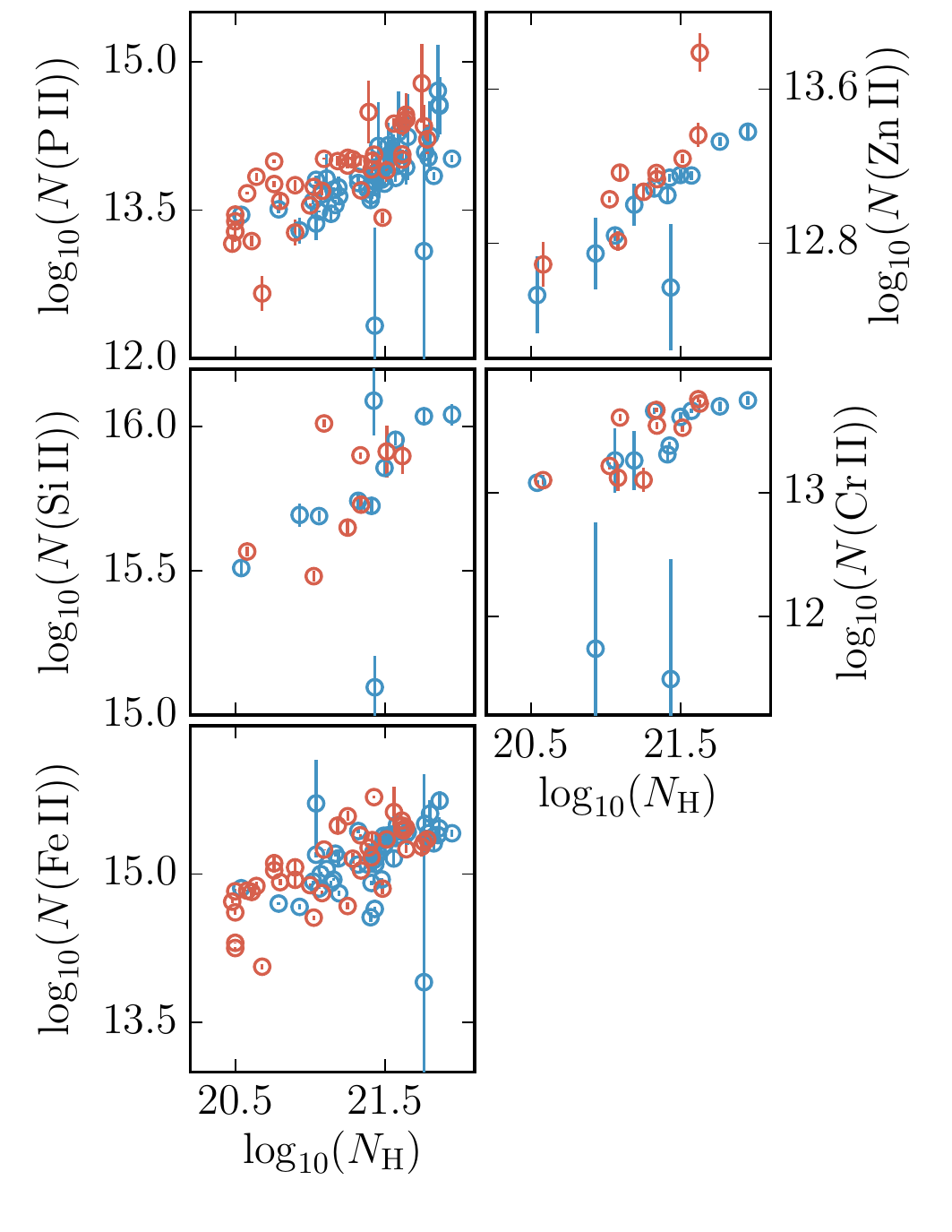}
  \caption{Metal ion column density plotted against the total column density
  of atomic and molecular hydrogen. LMC sightlines are denoted in red, SMC
  sightlines are in blue. Sightlines without COS data are shown
  without errorbars.}
  \label{fig:ioncolumns}
\end{figure}

  \begin{deluxetable*}{lccccccc}
  \tablecaption{Logarithmic ion column densities towards LMC and SMC targets in the COS sample}
  \tabletypesize{\scriptsize}
  \tablewidth{0pt}
  \tablehead{\colhead{Target} & \colhead{Galaxy} & \colhead{$\log_{10}(N(\mathrm{P\,II}))$} & \colhead{$\log_{10}(N(\mathrm{Zn\,II}))$} & \colhead{$\log_{10}(N(\mathrm{Si\,II}))$} & \colhead{$\log_{10}(N(\mathrm{Cr\,II}))$} &\colhead{$\log_{10}(N(\mathrm{Fe\,II}))$} & \colhead{$\log_{10}(N_\mathrm{H})$} \\ 
  \colhead{} & \colhead{} & \colhead{$[\mathrm{cm}^{-2}]$} & \colhead{$[\mathrm{cm}^{-2}]$} & \colhead{$[\mathrm{cm}^{-2}]$} & \colhead{$[\mathrm{cm}^{-2}]$} &\colhead{$[\mathrm{cm}^{-2}]$} & \colhead{$[\mathrm{cm}^{-2}]$}}
  \startdata
    AzV 327 & SMC &$13.30 \pm 0.13$ & $12.75 \pm 0.19$ & $15.69 \pm 0.04$ & $11.74 \pm 1.02$ & $14.67 \pm 0.02$ & $20.93$ \\
    SK 143 & SMC & $12.33 \pm 0.99$ & $12.57 \pm 0.33$ & $15.10 \pm 0.11$ & $11.49 \pm 0.97$ & $14.65 \pm 0.03$ & $21.43$ \\
    AzV 47 & SMC & $13.77 \pm 0.03$ & $13.08 \pm 0.03$ & $15.74 \pm 0.02$ & $13.66 \pm 0.03$ & $15.43 \pm 0.02$ & $21.32$ \\
    AzV 238 & SMC & $13.65 \pm 0.04$ & $13.05 \pm 0.04$ & $15.73 \pm 0.02$ & $13.31 \pm 0.07$ & $14.91 \pm 0.02$ & $21.41$ \\
    AzV 95 & SMC & $13.77 \pm 0.01$ & $13.15 \pm 0.03$ & $15.86 \pm 0.02$ & $13.61 \pm 0.04$ & $15.30 \pm 0.01$ & $21.50$ \\
    SK 116 & SMC & $13.83 \pm 0.04$ & $13.15 \pm 0.02$ & $15.95 \pm 0.02$ & $13.66 \pm 0.02$ & $15.36 \pm 0.03$ & $21.57$ \\
    SK 9 & SMC & $13.09 \pm 1.38$ & $13.33 \pm 0.02$ & $16.04 \pm 0.03$ & $13.70 \pm 0.04$ & $13.91 \pm 2.10$ & $21.76$ \\
    AzV 476 & SMC & $14.02 \pm 0.04$ & $13.38 \pm 0.04$ & $16.04 \pm 0.04$ & $13.75 \pm 0.04$ & $15.41 \pm 0.04$ & $21.95$\\
  SK-65 22 & LMC & $13.67 \pm 0.01$ & $12.69 \pm 0.12$ & $15.57 \pm 0.02$ & $13.10 \pm 0.04$ & $14.83 \pm 0.01$ & $20.58$ \\
  SK-67 5 & LMC & $13.73 \pm 0.02$ & $13.03 \pm 0.02$ & $15.48 \pm 0.02$ & $13.22 \pm 0.03$ & $14.56 \pm 0.02$ & $21.02$ \\
  SK-71 45 & LMC & $14.02 \pm 0.01$ & $13.17 \pm 0.04$ & $16.01 \pm 0.01$ & $13.61 \pm 0.03$ & $15.25 \pm 0.01$ & $21.09$ \\
  SK-66 172 & LMC & $13.95 \pm 0.03$ & $13.07 \pm 0.04$ & $15.65 \pm 0.02$ & $13.10 \pm 0.10$ & $14.68 \pm 0.04$ & $21.25$ \\
  SK-70 115 & LMC & $13.97 \pm 0.02$ & $13.16 \pm 0.03$ & $15.90 \pm 0.01$ & $13.67 \pm 0.02$ & $15.39 \pm 0.02$ & $21.34$ \\
  BI 173 & LMC & $13.70 \pm 0.02$ & $13.13 \pm 0.02$ & $15.73 \pm 0.02$ & $13.54 \pm 0.03$ & $15.04 \pm 0.02$ & $21.34$ \\
  BI 42 & LMC & $13.90 \pm 0.11$ & $13.24 \pm 0.02$ & $15.91 \pm 0.09$ & $13.52 \pm 0.04$ & $15.35 \pm 0.08$ & $21.51$ \\
  SK-68 135 & LMC & $14.06 \pm 0.05$ & $13.36 \pm 0.06$ & $15.90 \pm 0.06$ & $13.76 \pm 0.02$ & $15.45 \pm 0.03$ & $21.62$
  \enddata
  \end{deluxetable*}
  \label{tab:cos_columns}

  \begin{deluxetable}{lcccc}
    \tablecaption{Logarithmic ion column densities towards LMC and SMC targets in the FUSE sample}
    \tablehead{{Target} & {Galaxy} & {$\log_{10}(N(\mathrm{P\,II}))$} &
    {$\log_{10}(N(\mathrm{Fe\,II}))$} & {$\log_{10}(N_\mathrm{H})$} \\
     & & $[\mathrm{cm}^{-2}]$ & $[\mathrm{cm}^{-2}]$ & $[\mathrm{cm}^{-2}]$}
     \tablewidth{0pt}
     \startdata
    AzV 6 & SMC &$14.09 \pm 0.12$ & $15.40 \pm 0.05$ & $21.54$ \\
    SK 10 & SMC &$13.95 \pm 0.06$ & $15.50 \pm 0.02$ & $21.58$ \\
    SK 15 & SMC &$14.25 \pm 0.35$ & $15.61 \pm 0.13$ & $21.80$ \\
    SK 18 & SMC &$14.09 \pm 0.06$ & $15.51 \pm 0.04$ & $21.77$ \\
    SK 34 & SMC &$14.29 \pm 0.42$ & $15.39 \pm 0.05$ & $21.59$
    \enddata
  \tablecomments{Hydrogen column densities are from \citet{Welty:2012ct} and references therein. \\Table 3 is published in its entirety in the electronic edition of the Astrophysical Journal. A portion is shown here for guidance regarding its form and content.}
  \end{deluxetable}
  \label{tab:fuse_columns}

\subsection{Ion column densities}
\label{sec:specfit}
The analysis of an absorption spectrum can generally be divided into
two steps -- continuum fitting and column density recovery. 
In a typical continuum fitting procedure, one interpolates over
ISM absorption using neighboring parts of the spectrum as a guide. 
We do this interpolation using Gaussian process regression \citep{GPML}. 
A Gaussian process is a probability distribution over functions. 
Regression is a procedure for choosing a function that has a high
probability of having produced a set of observations. 
Gaussian process regression, then, is a procedure for choosing a function
that, on the one hand, has a high probability of having produced the
observed points and, on the other, has a high probability according to
the specified Gaussian process.
The procedure also produces an estimate of the function's pointwise
uncertainty. 
This estimate includes correlations between neighboring values.
Our chosen Gaussian process favors functions that resemble the
sum of a second-order polynomial and a smooth (i.e. infinitely differentiable), zero-mean perturbation.
For an example of what these continua look like over short wavelength
ranges, see the dashed black line in Figure \ref{fig:fit:sk116}.
For more details on this procedure, see Appendix \ref{ap:specfit}.

The next step, column density recovery, requires a procedure for
converting the observed amount of absorption into a column density or,
if different parts of the absorption are caused by different
absorbers, column densities. 
In our spectra, the absorption line corresponding to each transition
appears at several Doppler velocities, each corresponding to a different
line-of-sight object.
In Figure \ref{fig:fit:sk116}, each absorption line appears
near 0 km sec$^{-1}$, which corresponds to gas in the MW, and 130 km
sec$^{-1}$, which corresponds to gas in the SMC \citep{Brns:2005dl}.
Towards stars in the LMC, we see absorption at 0, 150, and 250 $\text{
  km sec}^{-1}$ due to gas in the MW, intervening high velocity clouds
\citep{Lehner:2009cg}, and and the LMC. 
These shifts cause velocity components of some absorption lines, in particular the Cr
II and Zn II lines near 2062 \AA\ and the three Fe II lines near 1143 \AA, to overlap. 
This can be seen in the second and third panels from the bottom of the
right column of Figure \ref{fig:fit:sk116} in which each absorption line
has two broad velocity components, but appears to have four.

To resolve the resulting confusion, we turn to a variation on Voigt
profile fitting. 
In standard Voigt profile fitting, one models the observed absorption
as coming from a series of coherent clouds.
Each cloud has a central velocity and (Gaussian) velocity width and contains some
amount of each species under investigation. 
One optimizes the central velocity, width, and amount of each species
in each cloud by generating absorption spectra, comparing them to the
observations, and minimizing the difference.
The number of clouds is chosen in order to get an acceptable fit according
to some criterion. 
We avoid choosing a set number of clouds and instead numerically
integrate out the dependence on the number of coherent clouds using a
transdimensional (i.e. operating over a parameter space of varying size) Markov
Chain Monte Carlo (MCMC) procedure \citep{Green:1995ut}.
This procedure allows us to include the effects of different possible
cloud decompositions in our quoted uncertainties.
Additional information about our Voigt profile fitting implementation can be
found in Appendix \ref{ap:specfit}.

We have used these two procedures to measure ion column
densities. These are plotted in Figure \ref{fig:ioncolumns}
and listed in Table \ref{tab:cos_columns} (COS sample) and Table
\ref{tab:fuse_columns} (FUSE-only sample).

\subsection{Elemental abundances and ionization corrections}
\label{sec:ioncorrections}
\begin{figure*}
  \centering
  \includegraphics[width=0.49\linewidth]{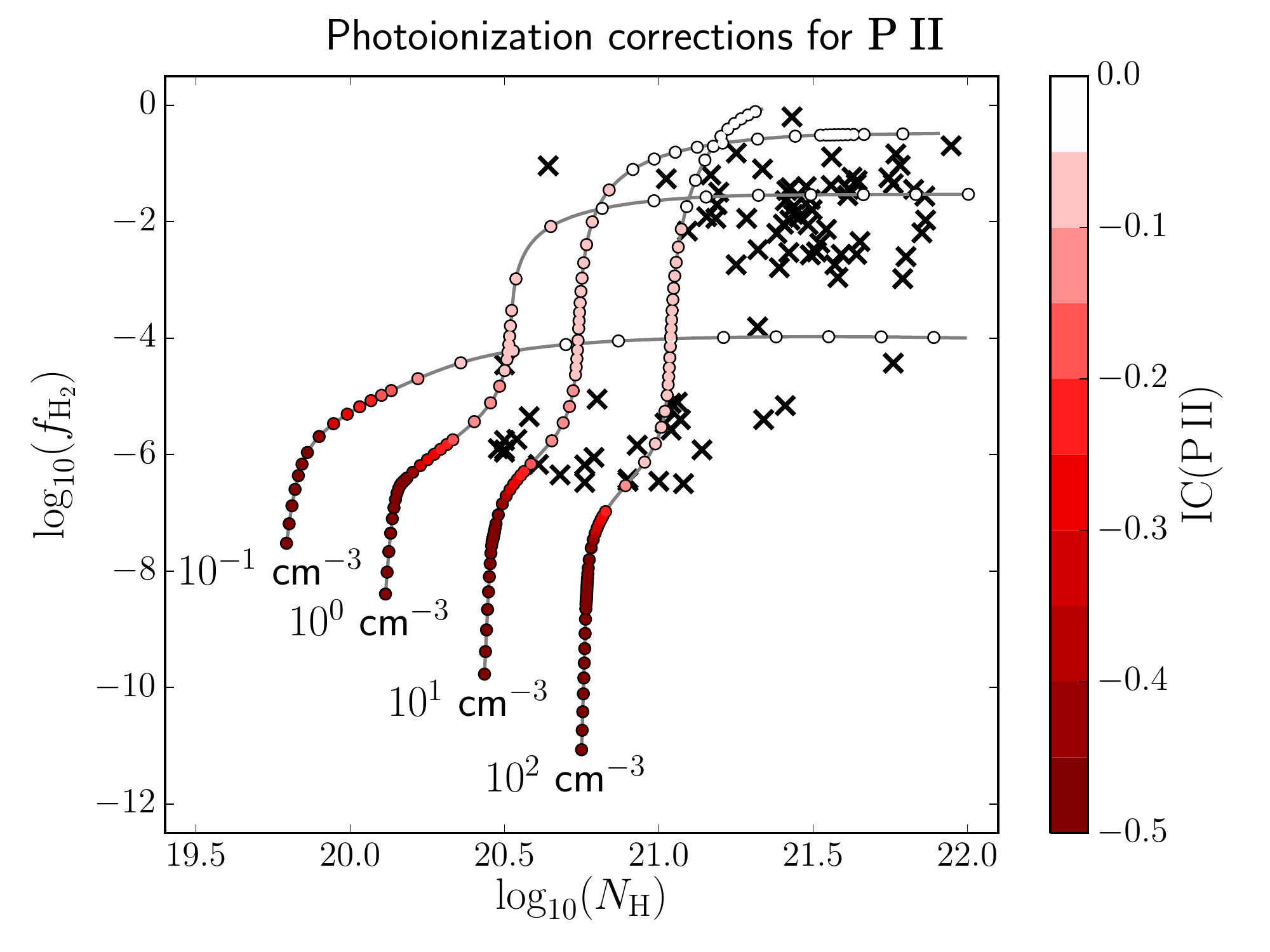}
  \includegraphics[width=0.49\linewidth]{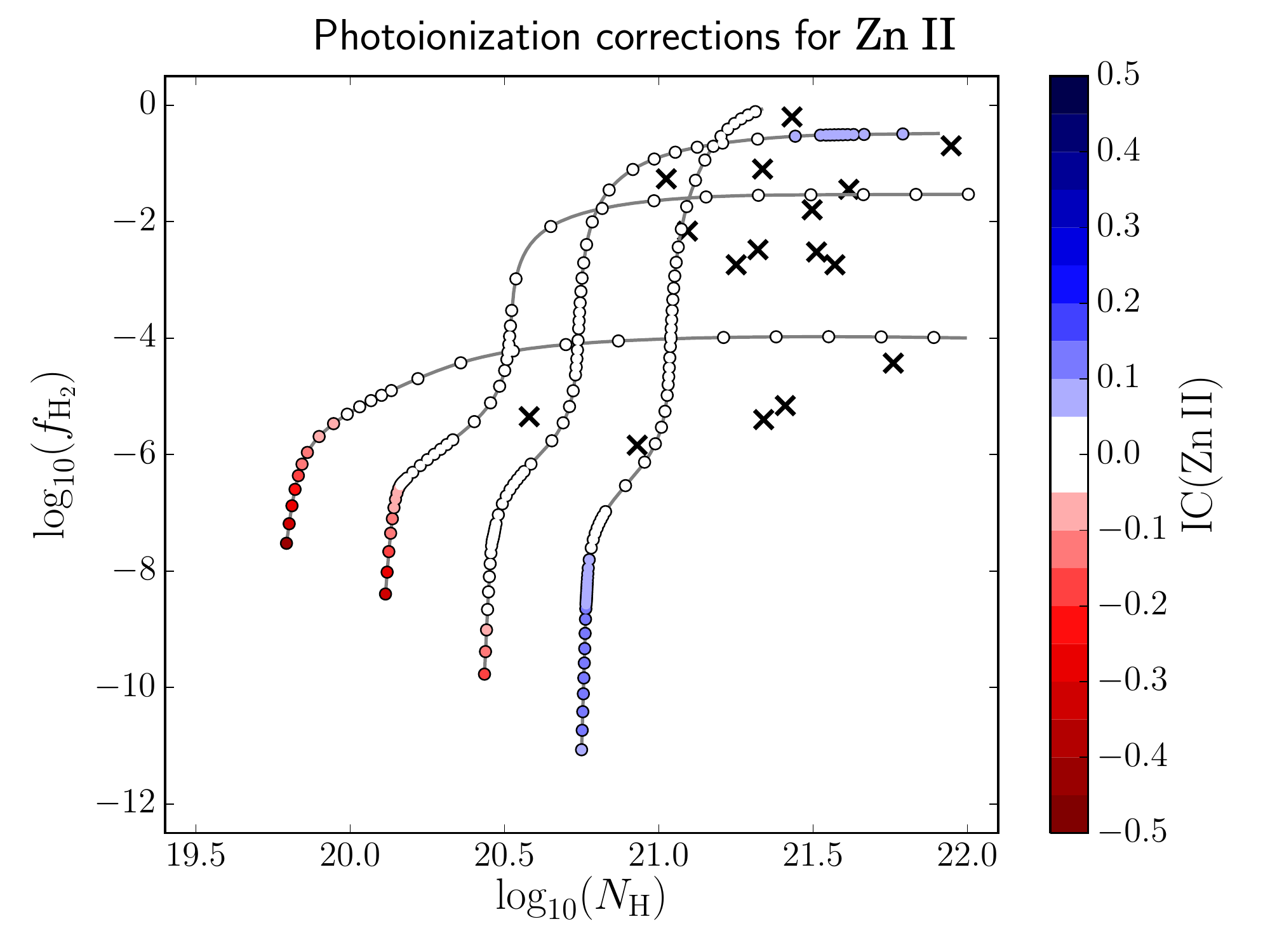}
  \caption{Theoretical ionization corrections (colors) for P II (left) and Zn II (right)
    as a function of the non-ionized hydrogen column density and the
    molecular hydrogen fraction. 
    Black points mark the hydrogen column densities and molecular
    hydrogen fractions along sightlines with FUSE (left panel) and COS
    (right panel) data. 
    Each track corresponds to a single
    total hydrogen volume density, which is marked at the track's base in left panel.
    These corrections would apply to a sightline towards an
    O2-type star in the LMC. 
    The corrections for every element in our study other than P II
    are similar to those of Zn II. 
    See Subsection \ref{sec:ioncorrections} for a 
    description of our photoionization calculations.}
  \label{fig:ioncorrections}
\end{figure*}

Next, we use the measured ion column densities to compute gas-phase elemental
abundances. 
From our ion measurements and literature atomic and molecular hydrogen
column densities \citep{Welty:2012ct}, we can directly compute gas-phase
ion abundances. 
Along a sightline with Zn II column density $N(\text{Zn\,II})$, atomic
hydrogen column density $N(\text{H\,I})$, and molecular hydrogen
column density $N(\text{H}_2)$, the gas-phase Zn II abundance
$\varepsilon_{gas}(\mathrm{Zn\,II})$ is
\begin{align}
  \varepsilon_{gas}(\mathrm{Zn\,II}) &= \log_{10} \left( \frac{N(\mathrm{Zn\,II})}{N(\mathrm{H\,I})+2N(\mathrm{H}_2)} \right)
\end{align}
Because this approximation does not account for every possible gas-phase form of
zinc and hydrogen, the Zn II abundance may be different from the true gas-phase Zn
abundance $\varepsilon_{gas}(\mathrm{Zn})$; the difference between the
ion and element abundances is called the ionization correction:
\begin{align}
  \text{IC}(\mathrm{Zn\,II}) &= \varepsilon_{gas}(\mathrm{Zn}) - \varepsilon_{gas}(\mathrm{Zn\,II}).
\end{align}
The magnitude of the ionization correction depends on the element's
ionization potentials and the ISM phases present along the line of sight \citep{2000ApJ...528..310S}.
Based on models of the ionization structure of clouds with H I and H$_2$
column densities similar to those measured along our sightlines, we
expect the ionization corrections to be negligible.

The ionization correction for the Zn II
abundance is small when, on the one hand, most of the Zn that is
spatially coincident with H I and H$_2$ is Zn II and, on the other hand,
most of the Zn that is not spatially coincident with H I and H$_2$ is
not Zn II. 
If the main ionizing source is the background star towards
which we are observing and its neighbors, then both conditions for a
small ionization correction will apply.
Both conditions require radiation fields that should affect a
qualitatively small fraction of the total gas.
Because the ionization potential from H I to H II is lower than
the ionization potential from Zn II to Zn III, the only way to miss
some of the spatially coincident Zn is to attenuate the radiation field above the
ionization potential from Zn I to Zn II. 
Since we can see non-negligible fluxes at and below the wavelength corresponding to this ionization
potential, 1320 \AA, the gas along the line of sight cannot contain
significant amounts of Zn I.
To violate the second condition, that Zn II be present without H I or
H$_2$, we need a radiation field that is capable of ionizing H I but
not Zn II. 
This requires photons with wavelengths between 690 \AA\, and 912
\AA, which corresponds to a fairly narrow shell within an H II
region. 
The ionization potentials to and from the other ions in this study
imply similar qualitative limits on the corresponding ionization corrections.

We can make these qualitative limits quantitative using
models of clouds illuminated by UV-bright stars. 
We do not know the volume density of the gas along each sightline, so
we have cloud models with constant volume densities of 0.1, 1, 10, and 100 particles per
$\text{cm}^3$. 
The spectral types of our background stars range from B0 to O2, so we
run each cloud model with a B0 star and with an O2 star in order to
bracket the range of possible ionization corrections; the 
stellar spectra come from \citet{2003ApJS..146..417L}. 
Stellar spectra, gas cooling, and gas shielding all depend on the
metallicity, so we run each star-cloud pair at the LMC and SMC
metallicities. 
These and other model parameters are listed and explained in Appendix \ref{ap:cloudy}.
For each combination of spectral type, volume density, and
metallicity, we use Cloudy \citep[version 13.02,][]{2013RMxAA..49..137F} to
calculate the cloud ionization structure.

From the output of these calculations, we derive ionization corrections
and molecular hydrogen fractions as a function of H I and H$_2$ column
density, where the molecular hydrogen fraction $f_{\mathrm{H}_2}$ is
\begin{align}
  f_{\mathrm{H}_2} &= \frac{2 N\left(\mathrm{H}_2\right)} {2 N\left(\mathrm{H}_2\right) + N\left( \mathrm{H\,I} \right)}.
\end{align}
Ionization corrections for P II and Zn II are shown in Figure
\ref{fig:ioncorrections}, along with the positions of our observations in the
hydrogen column-molecular fraction plane. 
These corrections are for an O2 star in the LMC.
Almost all of the stars towards which we observe have
later spectral types and, correspondingly, lower-magnitude P II corrections.
The P II ionization corrections for this specific combination of metallicity and illuminating star have larger absolute values than the ionization corrections for any other ion in any other model.
In all cases other than P II at the LMC metallicity illuminated by an
O2 star, the ionization corrections are similar in sign and magnitude to those of Zn II in Figure \ref{fig:ioncorrections}, whose largest magnitude in the part of the diagram occupied by our observations is less than 0.1 decades. 
To within our typical uncertainties, which are about 0.1 decades, our ion abundances should be equal
to the gas-phase elemental abundances.
\subsection{Depletions}
\label{sec:depletions}
\begin{figure}
  \centering
  \includegraphics[width=\linewidth]{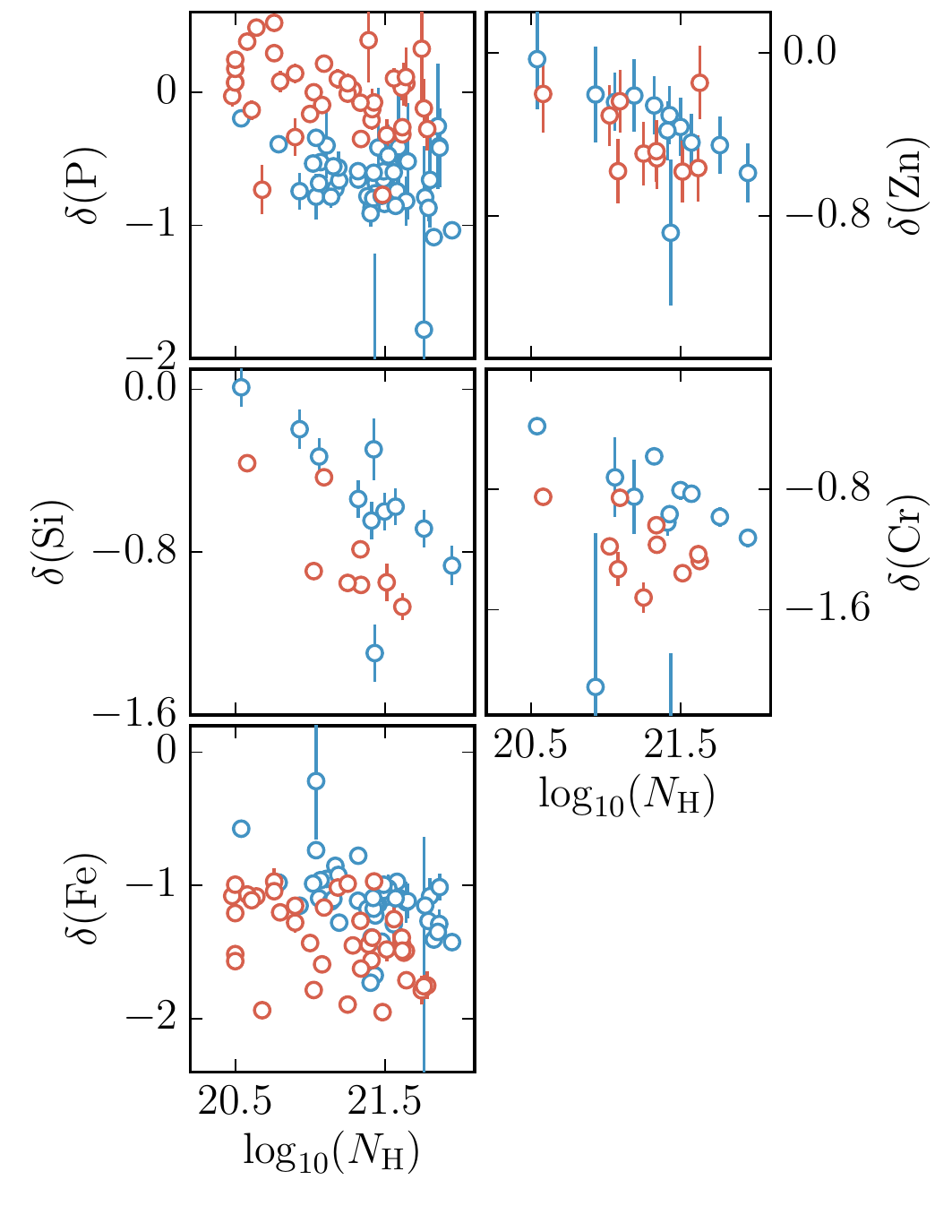}
  \caption{Elemental depletions relative to the ISM abundances of the
    LMC and SMC plotted against the total column density
  of atomic and molecular hydrogen. Depletions towards the LMC and SMC are shown in red and blue.}
  \label{fig:depletions}
\end{figure}

  \begin{deluxetable*}{c c c c p{2em}}
  \tablecaption{Photospheric abundances in the LMC and SMC }
    \tablehead{
      \colhead{Element} & \colhead{LMC abundance} & \colhead{SMC abundance} &
    \colhead{References} \\
     \colhead{} & \colhead{$[\mathrm{cm}^{-2}]$} & \colhead{$[\mathrm{cm}^{-2}]$} & \colhead{}
     }
     \tablewidth{0pt}
  \startdata
    C & $7.94 \pm 0.10 \pm 0.05$ & $7.52 \pm 0.10 \pm 0.04$ & 1, 2, 3, 5, 6, \\ & & & 7, 8, 10, 11 \\ 
    O & $8.50 \pm 0.11 \pm 0.03$ & $8.14 \pm 0.08 \pm 0.04$ & 1, 2, 3, 5, 6, \\ & & & 9, 10, 12 \\ 
    Mg & $7.26 \pm 0.08 \pm 0.04$ & $6.88 \pm 0.06 \pm 0.03$ & 1, 2, 4, 5, 6, \\ & & & 9, 10, 11, 12 \\ 
    Si & $7.35 \pm 0.10 \pm 0.02$ & $6.96 \pm 0.07 \pm 0.09$ & 1, 2, 4, 5, 6, \\ & & & 9, 10, 12\\ 
    P & 5.1 & 4.7 & 0 \\
    Cr & $5.37 \pm 0.07 \pm 0.03$ & $4.92 \pm 0.10 \pm 0.05$ & 1, 2,
    4, 9, 11, 12\\ 
    Fe & $7.32 \pm 0.08 \pm 0.03$ & $6.89 \pm 0.08 \pm 0.03$ & 1, 2,
    4, 6, 9, 11, 12 \\ 
    Ni & $5.92 \pm 0.07 \pm 0.03$ & $5.52 \pm 0.18 \pm 0.07$ & 1, 2,
    4, 9, 11 \\ 
    Zn & $4.31 \pm 0.15 \pm 0.15$ & $4.02 \pm 0.20 \pm 0.14$ & 8, 9,
                                                               11
    \enddata
    \tablecomments{Abundances are given in the form $\log_{10}(\mathrm{X}/\mathrm{H}) + 12$. 
    The first uncertainty is the standard deviation of the abundance value; the second is the standard deviation of the star-to-star dispersion.\\ 
    {\bf References.} (0) Solar $\times Z/Z_{\odot}$ \citep{2010Ap&SS.328..179G}, 
    (1) \citet{Chekhonadskikh:2012jy}, 
    (2) \citet{1995A&A...293..347H}
    (3) \citet{1997A&A...323..461H}
    (4) \citet{1997A&A...324..435H}
    (5) \citet{2007A&A...466..277H}
    (6) \citet{Korn:2002jv}
    (7) \citet{2005ApJ...633..899K}
    (8) \citet{1992ApJS...79..303L}
    (9) \citet{Luck:1998wz},
    (10) \citet{Rolleston:2003gu}, 
    (11) \citet{Russell:1989fs}
    (12) \citet{1999ApJ...518..405V}
    }
  \end{deluxetable*}
  \label{tab:abundances}

The depletion $\delta(\X)$ of an element $\X$ is the fraction of the element's
assumed reference, or gas- and solid-phase total, abundance $\epsilon_{\mathrm{ref}}$ that we
observe in the gas phase:
\begin{align}
  \delta(\X) &= \log_{10}\left( \frac{ \left(\frac{N(\X)}{N_\mathrm{H}}\right)_{\mathrm{gas}}} {
  \left(\frac{N(\X)}{N_\mathrm{H}}\right)_{\mathrm{ref}}}\right)\\
  &= \epsilon(\X)_{\mathrm{gas}} - \epsilon(\X)_{\mathrm{ref}},
\end{align}
where $\epsilon_{\mathrm{gas}}$ is the observed gas-phase abundance. 
The more negative $\delta$ is, the more of the element is missing,
presumably in the solid phase. This makes $\delta$ a proxy
for the gas-to-dust ratio.

In order to compute depletions, we need reference abundances.
The composition of a star's photosphere is a superposition of the ISM
composition at the star's formation time with the effects of various
enrichment processes. This makes recently-formed stars that have not
yet undergone self-enrichment good proxies for the present-day ISM
composition.

There have been a number of spectroscopic measurements of the
composition of luminous stars in both MCs. 
Because no individual study includes all of the elements we are
interested in, we have to combine their measurements.
We believe that we cannot simply average measurements across the
studies because of significant between-study differences, repeat
observations and analyses of the same stars by multiple studies, and
missing uncertainty information.
Instead, we partially pool measurements across studies using a
multilevel linear model \citep{BrowneDraper2006}. 
The model is described in detail in Appendix \ref{ap:meta}.
Our adopted reference abundances, abundance uncertainties, and
estimated intrinsic (i.e. location-to-location) abundance variances
are listed in Table \ref{tab:abundances}. Figure \ref{fig:depletions}
shows the depletions we obtain by assuming these reference abundances.

In the next section, we analyze how depletions vary from sightline to
sightline. 

\subsection{Linear depletion relations}
\label{sec:depletiondist}
\begin{figure*}
  \centering
  \includegraphics[width=0.95\linewidth]{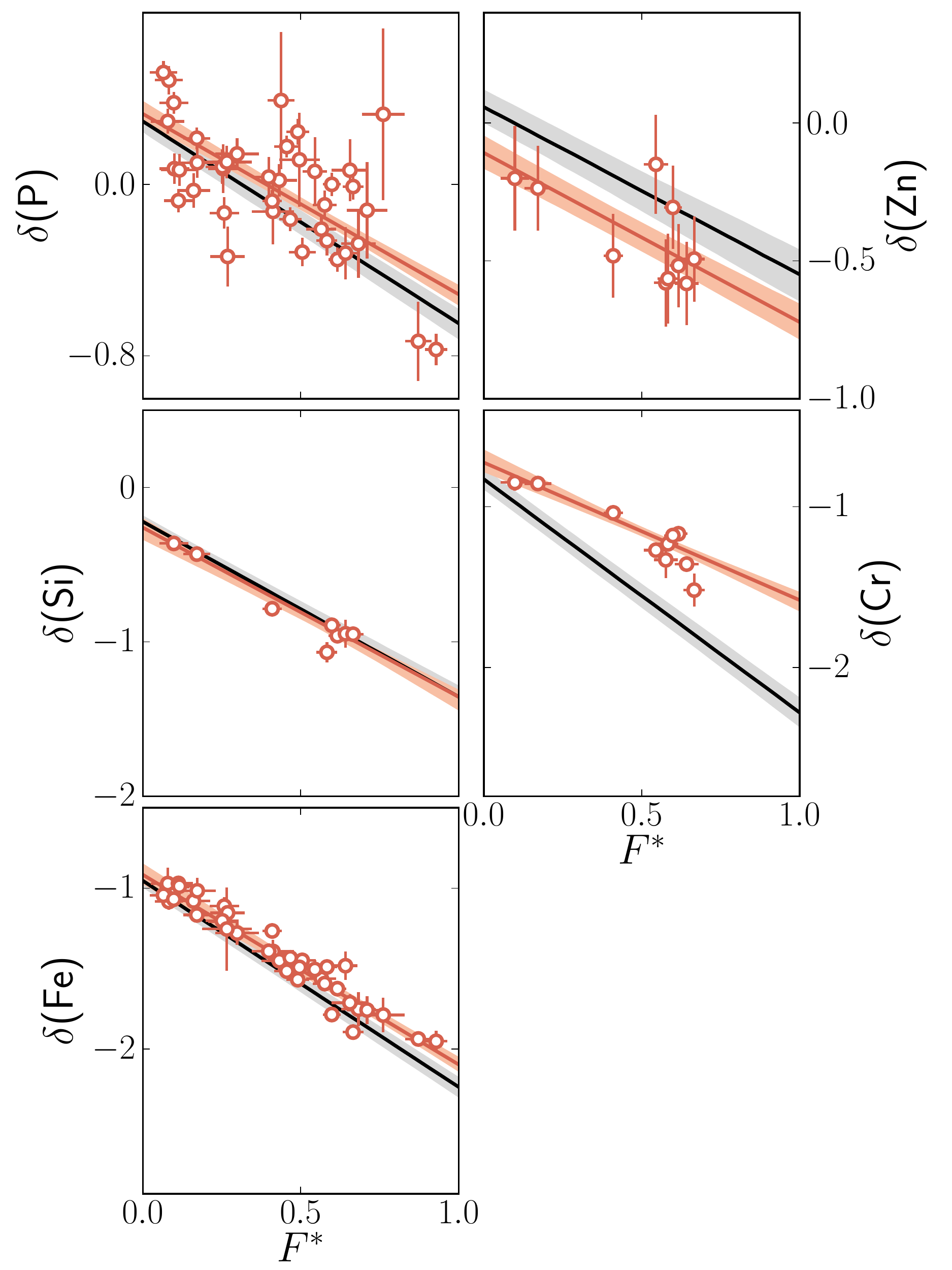}
  \caption{Fits of the Jenkins parametrization to elemental depletions
  in the LMC ISM. Data points include our observations, our analysis
  of archival FUSE data, and measurements from \citet{1997ApJ...474L..95R}.
  Shaded regions show $68\%$ credible
  intervals about the median fit. The black line is the MW best fit
  from \citet{Jenkins:2009ke}.}
  \label{fig:lmc:jfits}
\end{figure*}
\begin{figure*}
  \centering
  \includegraphics[width=0.95\linewidth]{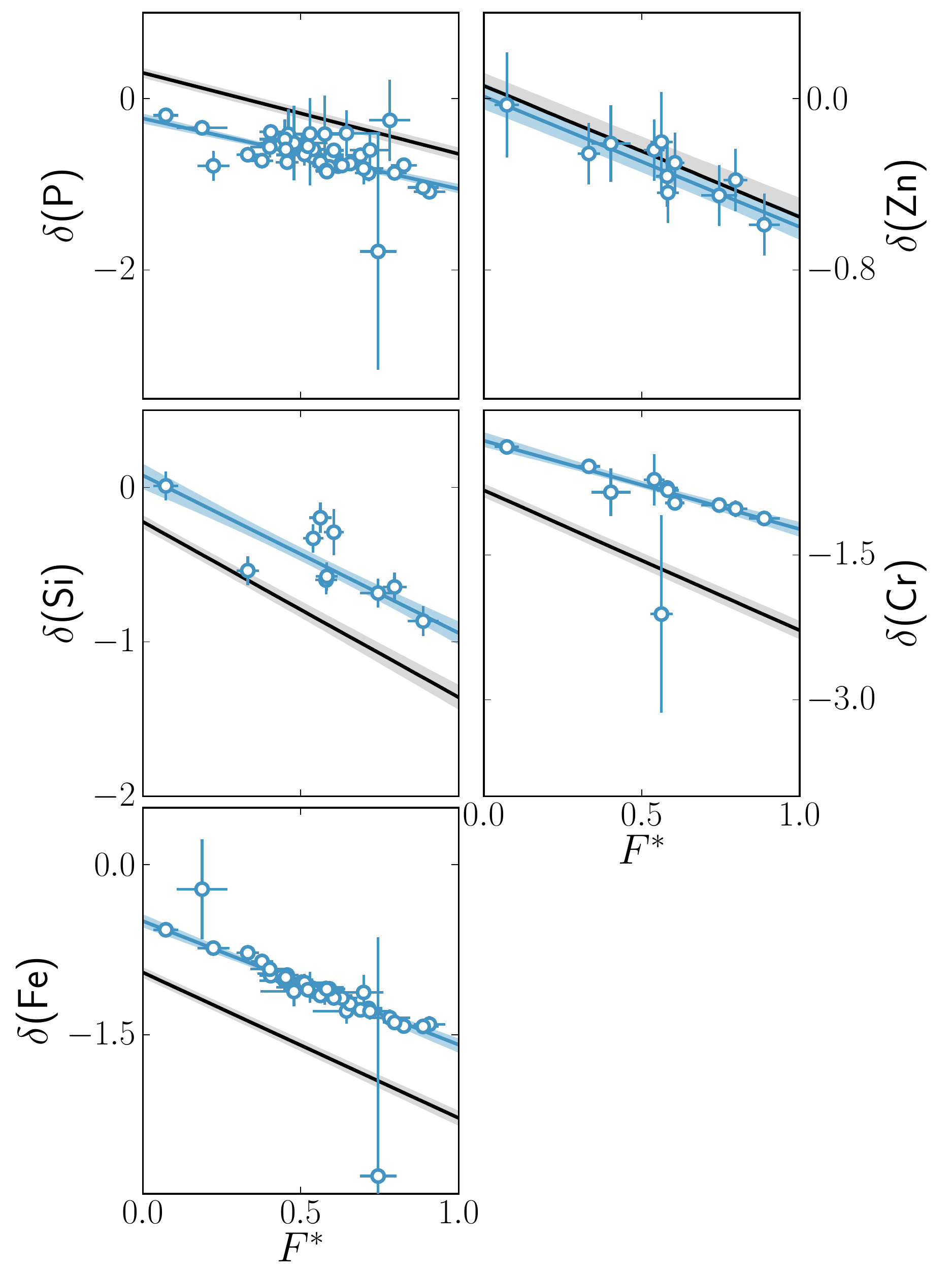}
  \caption{Fits of the Jenkins parametrization to elemental depletions
  in the SMC ISM. Data points include our observations, our analysis
  of archival FUSE data, and measurements from
  \citet{1997ApJ...474L..95R}, \citet{2001ApJ...554L..75W}, and \citet{2006ApJ...636..753S}.
  Shaded regions show $68\%$ credible
  intervals about the median fit. The black line is the MW best fit
  from \citet{Jenkins:2009ke}.}
  \label{fig:smc:jfits}
\end{figure*}

\begin{table}
  \centering
  \caption{Abundance uncertainties and depletion parameters}
  \begin{tabular}{rcc}
   \hline &{\bf {LMC}}& {\bf {SMC}}\\
    \hline
    {\bf {P:}}\hspace{0.1in} $\sigma_\epsilon$ & \ldots & \ldots\\
    $\delta(\mathrm{P})_0$ & $0.30 \pm 0.03$ & $-0.30 \pm 0.04$ \\
    $A_\mathrm{P}$ & $-0.71 \pm 0.07$ & $-0.83 \pm 0.07$\\

    {\bf {Zn:}}\hspace{0.1in} $\sigma_\epsilon$ & 0.15 & 0.2 \\
    $\delta(\mathrm{Zn})_0$ & $ -0.08 \pm 0.06$ & $-0.04 \pm 0.06$ \\
    $A_\mathrm{Zn}$ & $-0.60 \pm 0.06$ & $-0.61 \pm 0.06$\\

    {\bf {Si:}}\hspace{0.1in} $\sigma_\epsilon$ & 0.1 & 0.07 \\
    $\delta(\mathrm{Si})_0$ & $-0.32 \pm 0.03$ & $-0.01 \pm 0.08$ \\
    $A_\mathrm{Si}$ & $-0.92 \pm 0.08$ & $-1.03 \pm 0.15$\\

    {\bf {Cr:}}\hspace{0.1in} $\sigma_\epsilon$ & 0.07 & 0.1 \\
    $\delta(\mathrm{Cr})_0$ & $-0.77 \pm 0.04$ & $-0.36 \pm 0.06$ \\
    $A_\mathrm{Cr}$ & $-0.71 \pm 0.09$ & $-0.97 \pm 0.11$\\

    {\bf {Fe:}}\hspace{0.1in} $\sigma_\epsilon$ & 0.08 & 0.08 \\
    $\delta(\mathrm{Fe})_0$ & $-0.96 \pm 0.02$ & $-0.57 \pm 0.05$ \\
    $A_\mathrm{Fe}$ & $-1.00 \pm 0.06$ & $-1.13 \pm 0.07$\\
    \hline
  \end{tabular}\\
    {\bf Notes.} We describe the
    abundance uncertainties $\sigma_\epsilon$ in
    Section~\ref{sec:depletions} and the depletion parameters
    $\delta(\mathrm{X})_0$ and $A_\mathrm{X}$ in Section~\ref{sec:depletiondist}.
  \label{tab:jenk:fit}
\end{table}
We would like to use the depletions derived in the previous section to
estimate ranges of solid-phase elemental abundances in the MCs and
compare them with abundances in the MW. For these estimates to be
robust, we need a representative sample of each MC's diffuse neutral
medium. The difference between the lowest and highest depletions of Fe
and P, which we measure using FUSE spectra, is approximately equal to
the same difference in the MW, suggesting that the FUSE sample is
representative of the same range of diffuse neutral medium conditions
as is covered by MW datasets.

In order to make the same depletion range estimates for the elements
that are only available in the (unrepresentative) COS sample, we need
a way of combining its elemental information with the FUSE
sample's population information. For this purpose, we use the Jenkins depletion
parametrization \citep{Jenkins:2009ke}. In this parametrization, the observed
depletion $\delta(X)_i$ of element $\X$ along a sightline $i$ is
given by 
\begin{align}
  \delta(\X)_i &= \delta(\X)_0 + A_\X  F^*_i,
\end{align}
where $\delta(\X)_0$ and $A_\X$ are the minimum and
range of depletions of element $\X$ in the diffuse neutral
medium, and $F^*_i$ is a parameter representing the overall depletion
level of sightline $i$.
$F^*_i$ is kept fixed for all elements along
sightline $i$ and $\delta(\X)_0$ and $A_\X$  are kept
fixed for an element $\X$ across all sightlines.

This model is underdetermined; we can rescale or shift combinations of
parameters without changing the obserables $\delta(\X)_i$.
We can scale a set of slope parameters $A_\X$ by a constant $C$ and
scale the $F_i^*$ values by $1/C$.
At fixed $A_\X$, we can add a constant $C$ to
$\delta(\X)_0$ and subtract a factor of $C/A_\X$ from the $F_i^*$.
We fix these two degeneracies and connect our $F_i^*$ values with
those of \citet{Jenkins:2009ke} by imposing the MW values as priors on
$\delta(\mathrm{Zn})$ and $A_\mathrm{Zn}$.
We chose this prior because its implied $F^*$ zero point location has a clear 
interpretation -- $F^*$ is zero when the depletion of Zn is (up to
uncertainty) zero. 
Since Zn is a volatile element in the MW, meaning that its minimum
depletion in the DNM is zero, we expect it to continue to be
volatile in more dust- and metal-poor galaxies such as the MCs.

Our priors on the remaining parameters are uninformative.
We impose broad uniform priors on the other depletion zero points and
slopes, a uniform prior with bounds $F^*_\mathrm{min}$ and
$F^*_\mathrm{max}$ on the $F_i^*$, and uniform priors with bounds
$(-3, 0.5)$ and $(0.5, 3)$ on, respectively, $F^*_\mathrm{min}$ and
$F^*_\mathrm{max}$. 
The $F^*$ bounds are included as parameters because we do not \emph{a
  priori} know how much of the possible $F^*$ range our observations will
fill. 
For both MCs, the $F^*$ ranges ended up being approximately $(0,1)$.

We implemented and generated samples from this model using the
Bayesian statistical analysis module PyMC \citep{PyMC}. 
Our fits to the LMC and SMC depletions are shown in Figures
\ref{fig:lmc:jfits} and \ref{fig:smc:jfits}. 
For each element, the reference abundance uncertainties, depletion zero points $\delta(\X)_0$, and depletion slopes $A_\X$ are listed in Table~\ref{tab:jenk:fit}.
The depletion zero points of the refractory elements Si, Cr, and Fe are significantly less negative in the SMC than they are in MW or LMC. 
The minimum Si depletion is particularly notable, since it is consistent with 0.
However, if we include the uncertainty of the reference Si abundance, the minimum Si depletion is also within $1\sigma$ of 0.13 in logarithmic units, which corresponds to 1/4 of the Si being out of the gas phase.
The depletion slope of Cr in the LMC and SMC is similar and signficantly smaller than in the MW.
The other depletion slopes do not change significantly between the three galaxies.
In the common interpretation of depletion variations within a galaxy, depletion zero points are associated with the composition of the refractory dust cores while depletion slopes are associated with the rate at which dust mantles either grow or are destroyed. 
By this interpretation, the fractional gas-to-solid transfer rate is roughly constant between galaxies while the refractory core composition changes. 
We will discuss the
implications of these zero point and slope changes further in
Sections~\ref{sec:evo:abund} and \ref{sec:evo:dla}.

\section{Quantitative analysis: solid-phase}
\label{sec:qa:dust}
In this section, we describe our derivation of solid-phase abundances
and GDRs from measured gas-phase and assumed total ISM abundances. 
\subsection{Solid-phase abundances}
We can reasonably assume that all of the metals that are missing from the gas-phase are in dust and can use this assumption to derive a
solid-phase abundance $\varepsilon(\X)_\mathrm{dust}$:
\begin{align}
  \varepsilon(\X)_\mathrm{dust} &= \log_{10} \left(
    10^{\varepsilon(\X)_\mathrm{ref}} - 10^{\varepsilon(\X)_\mathrm{gas}} \right)
    \label{eqn:gas:to:dust:abund}\\
    &= \log_{10} \left( 10^{\varepsilon(X)_\mathrm{ref}} \left( 1 - 10^{\delta(X)}\right) \right),
    \label{eqn:dep:to:dust:abund}
\end{align}
where $\varepsilon(\X)_\mathrm{ref}$ is the reference (i.e. theoretical gas- and solid-phase total) abundance, $\varepsilon(\X)_\mathrm{gas}$ is the gas-phase abundance, and $\delta(\X)$ is the depletion of the gas-phase abundance relative to the reference abundance. 
By combining the depletion-based formulation of Equation~\ref{eqn:dep:to:dust:abund} with the depletion parametrization described in Section~\ref{sec:depletiondist}, we can get a continuous empirical relation between $F^*$ and the solid-phase abundance of an element. We explore solid-phase abundances and related quantities further in Sections~\ref{sec:gdr} and \ref{sec:FeSi}.

\subsection{GDRs}
\label{sec:gdr}
\begin{figure}
  \centering
  \includegraphics[width=\linewidth]{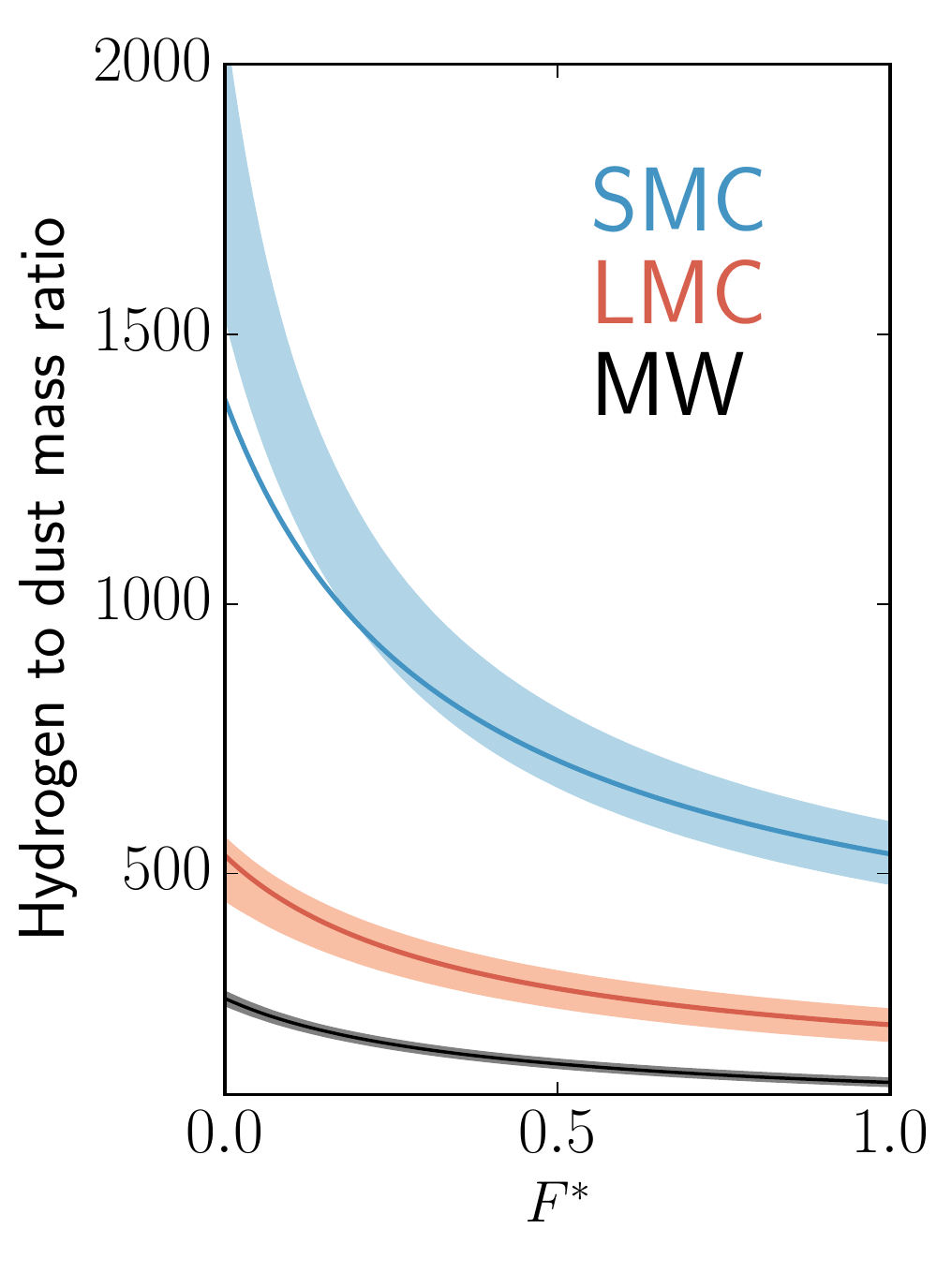}
  \caption{We show the surface mass GDR as a function of the over-all depletion
    level $F^*$, which increases from low depletion in all elements ($F^*=0$) to high depletion in all elements ($F^*=1$). The MW, LMC, and SMC are denoted by black, red, and blue. 
    The solid lines are derived by applying Equation~\ref{eq:dep:to:gdr} to each galaxy's best-fit reference abundances and the MW's best-fit depletion relations \citep{Jenkins:2009ke}. 
    The shaded regions are derived by applying Equation~\ref{eq:dep:to:gdr} to each galaxy's reference abundances and depletion relations, both with uncertainties. 
    The vertical extent of the shaded regions represent the 84\% posterior credible interval of the GDR at each value of $F^*$.}
  \label{fig:gdr}
\end{figure}

We can combine all of the elements' solid-phase abundances to compute a gas-to-dust mass ratio:
\begin{align}
  \frac{1}{\text{GDR}} = \frac{\Sigma_\mathrm{dust}}{\Sigma_\mathrm{gas}} &\approx
  \frac{1}{m_\mathrm{H}}\sum_X m_\X 10^{\varepsilon(X)_\mathrm{dust}},
  \label{eq:eps:to:gdr}
\end{align}
where $\X$ ranges over all elements and $m_\X$ is the atomic mass of element $\X$. 
This expression can be rewritten in terms of depletions $\delta(X)$:
\begin{align}
  \frac{\Sigma_\mathrm{dust}}{\Sigma_\mathrm{gas}} &\approx
  \frac{1}{m_\mathrm{H}}\sum_\X m_\X 10^{\varepsilon(X)_\mathrm{ref}} \left( 1 - 10^{\delta(\X)}\right).
  \label{eq:dep:to:gdr}
\end{align}
In the MW, the only elements that contribute significantly to this sum are C, O, Mg, Si, Fe, and Ni. 
All other elements have low reference abundances and/or negligible depletions. 
If we assume that the composition of dust in the MCs is even remotely like that of dust in the MW, we can restrict the sum in Equation \ref{eq:eps:to:gdr} to those elements without missing much of the total dust mass.

Since most of our sightlines do not have Si measurements and all of our sightlines do not have C, O, Mg, or Ni measurements, we compute GDRs using the linear depletion relations from Section \ref{sec:depletiondist}. 
We use our LMC and SMC relations for Si and Fe and the MW relations from \citet{Jenkins:2009ke} for C, O, Mg and Ni.
GDRs derived from these combination of depletion relations are shown as red (LMC) and blue (SMC) shaded regions in Figure \ref{fig:gdr}. 
For comparison, we show the MW's GDR in black and GDRs derived by assuming MW depletions at LMC and SMC reference abundances as solid red and blue lines.
We will refer to these GDRs as abundance-scaled MW GDRs to differentiate them from the first set, which we will refer to as \emph{the} LMC and SMC GDRs.

We can make three main observations about this figure.
First, regardless of which set of depletions we adopt, the GDR of all three galaxies changes by a factor of 2 from $F^*=0$ to $F^*=1$. 
This suggests that there is a significant amount of dust destruction and/or growth in the DNM of the LMC, SMC, and MW. 
Second, the GDRs of the LMC and the high-depletion part of the SMC are approximately the same as the corresponding abundance-scaled MW GDRs. 
Third, the GDR of the low-depletion part of the SMC is higher than the SMC-abundance-scaled MW GDR. 
The second observation may be an artifact of our assumptions, as the MC depletion relations that we do observe are always either as depleted or less depleted than the corresponding MW relations.
This suggests that the depletions we are assuming for some or all of C, O, Mg, and Ni are too high and implies that even the GDRs which assume MC depletions for Si and Fe are most likely underestimates. 
The third observation suggests that the \emph{dust-to-metals ratio} (DMR) decreases with decreasing metallicity.

At and above the metallicity of the SMC, the (galaxy-averaged) DMR is constant while below the SMC metallicity, the DMR ratio decreases with decreasing metallicity \citep{2014A&A...563A..31R}. 
From the first observation, we conclude that the DMR decreases by a factor of at least 2 within a galaxy. 
At medium to high over-all depletion (i.e. at $F^*=1$ or $0.5$), the DMR does not change from the MW to the LMC to the SMC. 
However, at low over-all depletion (i.e. at $F^*=0$), the DMR is constant from the MW to the LMC but decreases from the LMC to the SMC.
This suggests that in diffuse, low-depletion gas, the DMR stops being constant at a higher metallicity than in more dense, higher-depletion gas.

\subsection{The solid-phase stoichiometry of silicon and iron}
\label{sec:FeSi}

\begin{figure}
  \centering
  \includegraphics[width=\linewidth]{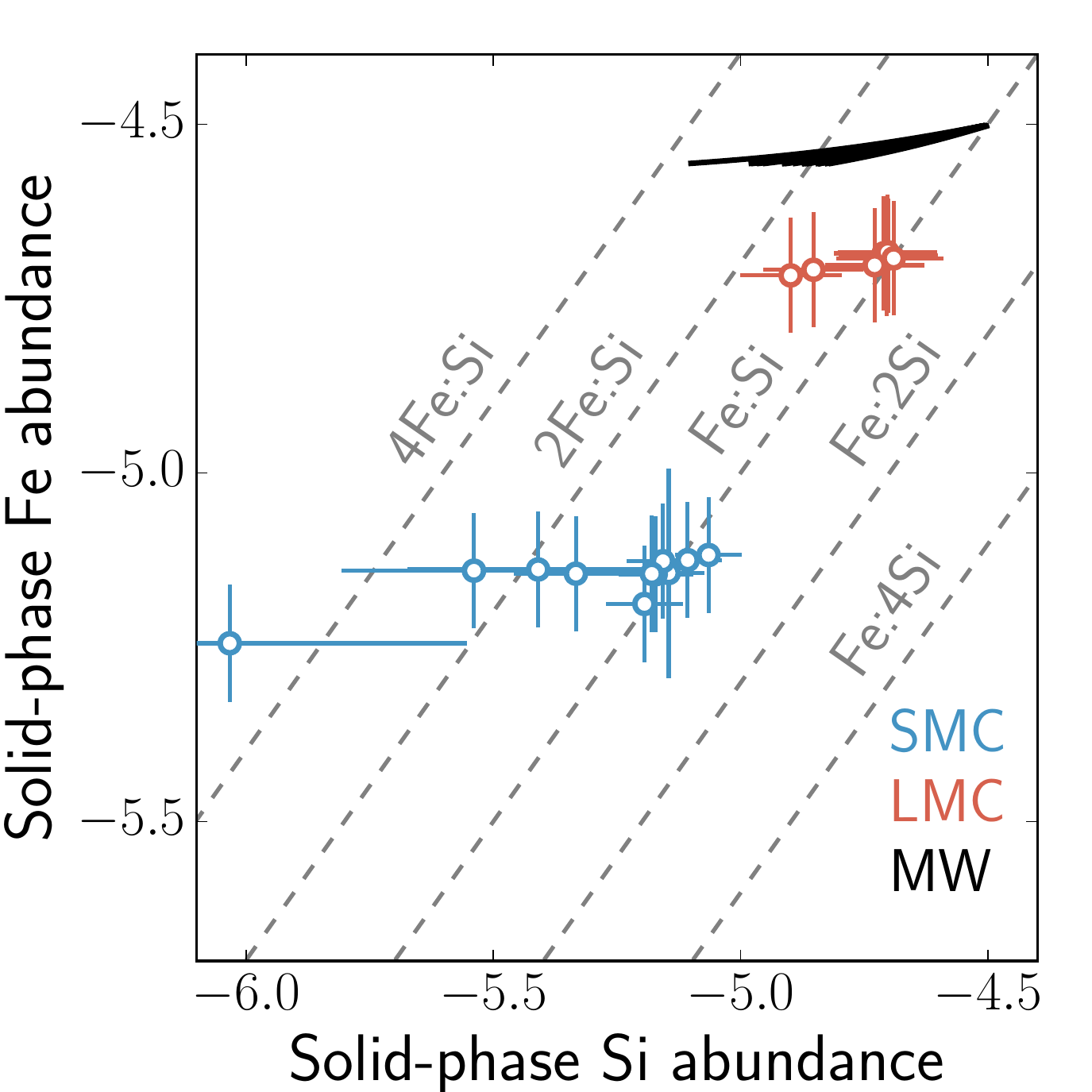}
  \caption{
    We show the solid-phase Fe abundance as a function of the solid-phase Si abundance.
    Data points from the SMC (blue) and LMC (red) are derived from reference abundances for the appropriate galaxy and gas-phase abundances using Equation~\ref{eqn:gas:to:dust:abund}, and data point error bars include uncertainties in both quantities. 
    The MW region (black) is derived from solar abundances and MW depletions \citep{Jenkins:2009ke} using Equation~\ref{eqn:dep:to:dust:abund}.
    The dashed grey lines mark levels of constant
    stoichiometric Fe to Si ratio.
    Within each galaxy, the over-all depletion level increases from left to right.
  }
  \label{fig:fesi}
\end{figure}

As shown in the previous section, the GDR is not constant throughout the DNM of each galaxy.
We now use depletion measurements to show that throughout each galaxy's DNM, the dust stoichiometry, or its relative bulk composition, also varies.
Using Equation~\ref{eqn:gas:to:dust:abund}, we have computed the solid-phase abundance of Si and Fe towards each sightline for which gas-phase column density measurements are available. 
These solid-phase abundances are shown in Figure~\ref{fig:fesi}, along with solid-phase abundances corresponding to the observed range of Si and Fe depletions in the MW \citep{Jenkins:2009ke}. 
The solid-phase Fe abundance is approximately constant across each galaxy, while the solid-phase Si abundance changes from the least- to the most-depleted sightlines by a factor of about 2. 

In all three galaxies, the Fe:Si abundance ratio in dust ranges from about 2:1 along the least-depleted sightlines to about 1:1 along the most-depleted sightlines. 
Since we are merely measuring the bulk dust composition, we cannot use these ratios to determine the composition of each dust subspecies. 
Nevertheless, the constant Fe abundance and variable Si abundance suggest that the process by which most of the Fe forms into dust is different from the process by which at least half of the Si forms into dust. 
This suggests that the Fe-rich dust component is either particularly resilient, if most of the abundance variations are due to dust destruction, has a faster growth rate, if most of the abundance variations are due to ISM dust growth, or some combination of the two. 
While this last fact has been known for MW dust for quite some time \citep[see, e.g.][]{1996ARA&A..34..279S}, we can now also confirm that it holds in other galaxies.

\section{Discussion}
\label{sec:discussion}
\subsection{The ISM metallicity in the Magellanic Clouds}
\begin{figure}
  \centering
  \includegraphics[width=\linewidth]{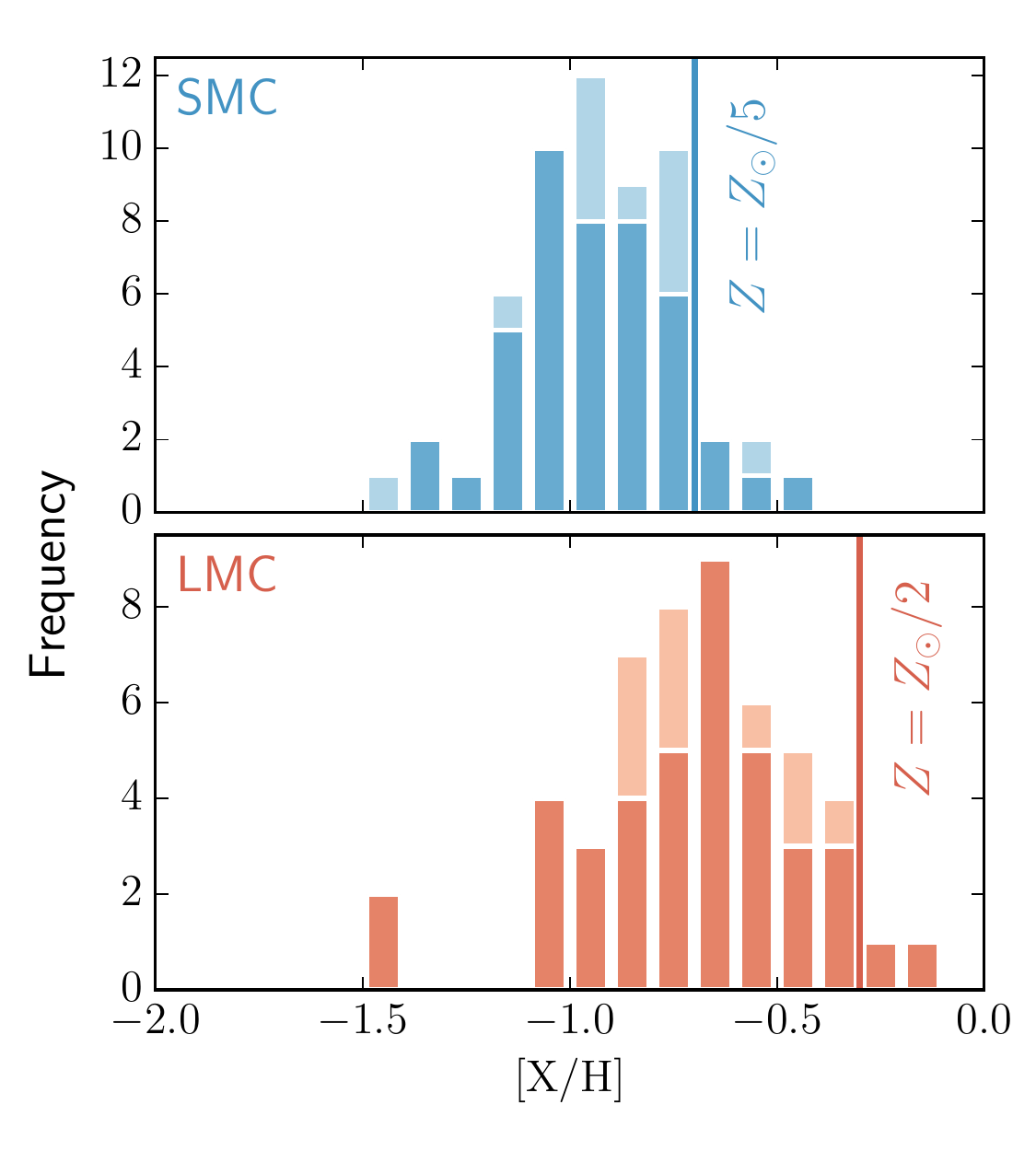}
  \caption{Bars show the distribution of measured
    gas-phase P (dark color) and Zn (light color) abundances relative to each element's solar abundance; vertical lines show the typically assumed
    metallicities of the LMC and SMC. P and Zn are relatively volatile elements. Their
    abundance at minimum depletion is a proxy for the ISM metallicity.
    Abundances greater than the mean MC metallicities (vertical lines) would indicate small-scale metallicity enhancements. 
    We see no evidence for localized metal enrichment along our sightlines.}
  \label{fig:metallicity}
\end{figure}
Studies of field star abundances in the MCs
(e.g. \citealt{Carrera:2008eh, Cioni:2009db}) have found mild
metallicity gradients as a function of galactocentric
distance. The metallicities of stars of similar ages but different
galactocentric radii tend to be the same, suggesting that these
gradients are due to star formation occuring at different
times in different locations. If the observed gradients were instead due
to ISM inhomogeneities, stars that formed at the same time
in different locations would have different metallicities.

We can use the abundances of our more volatile elements (P and Zn) to
estimate the amount of localized metal enrichment towards some of our
sightlines. At the lowest depletion levels, the gas-phase abundances of
P and Zn should approach their total ISM abundances. An undepleted metal-enriched
sightline would have a (local) metallicity greater than its
corresponding galactic mean. Figure \ref{fig:metallicity} shows the
distribution of P and Zn abundances in the MCs relative to the solar
abundance. 
While there are six apparently P-enriched and one apparently Zn-enriched sightlines across the pair of galaxies, this does not 
We do not find conclusive evidence of localized metal enrichment.

Since even the relatively volatile elements in this study can be
significantly depleted, we cannot exclude the possibility that some of
our sightlines are metal-poor. Our maximum measured gas-phase P and Zn abundances are consistent
with the typically assumed MC metallicities of $1/2$ and $1/5$ times
the solar metallicity for the LMC and SMC, respectively.

\subsection{Previous depletion studies of the Magellanic Clouds}
\label{sec:prevstudies}
\begin{figure}
  \centering
  \includegraphics[width=\linewidth]{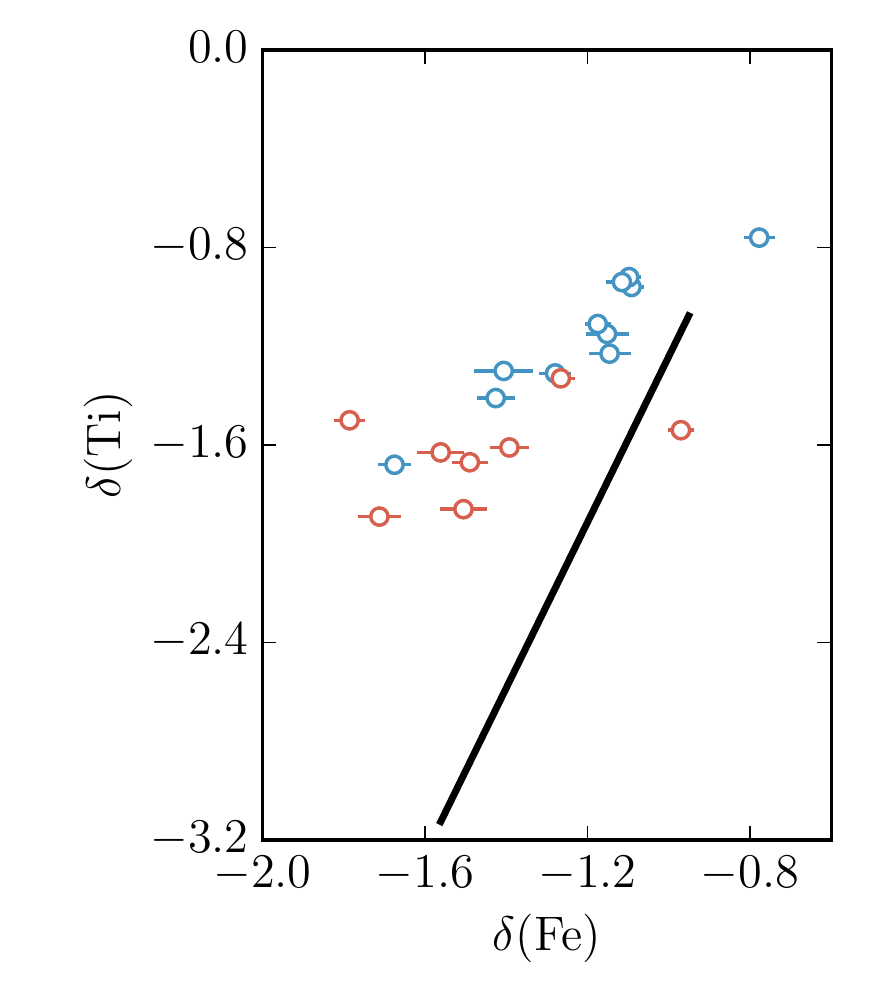}
  \caption{Fe (from this work) and Ti \citep{2010MNRAS.404.1321W} depletions in the LMC (red) and SMC (blue). For comparison, the linear depletion relation for Fe and Ti in the MW from \citet{Jenkins:2009ke} is shown in black.}
  \label{fig:ZnTi}
\end{figure}
Depletion studies of the MCs have been either medium-sized ($<20$ depletion measurements) surveys of single
elements or multi-element case studies of single sightlines. 
All of the targets in the multi-element case studies are part of our FUSE sample, and have been included in our analyses as described in Section~\ref{sec:fusetargets}.

\citet{2010MNRAS.404.1321W} presented an optical survey of
titanium (Ti) towards approximately twenty stars in either MC. 
Figure \ref{fig:ZnTi} shows a comparison of our Fe depletions and
their Ti depletions for the sightlines
which are included in both surveys. LMC and SMC sightlines are shown
in red and blue. The black line represents the typical correlation of
Fe and Ti depletions in the MW \citep{Jenkins:2009ke}. 
Across both galaxies, all but two of the targets are above this line, meaning that at fixed Fe depletion, MW sightlines tend to contain a smaller fraction of gas-phase Ti than MC sightlines.

\subsection{The evolution of gas-phase abundances with metallicity}
\label{sec:evo:abund}
\begin{figure}
  \centering
  \includegraphics[width=\linewidth]{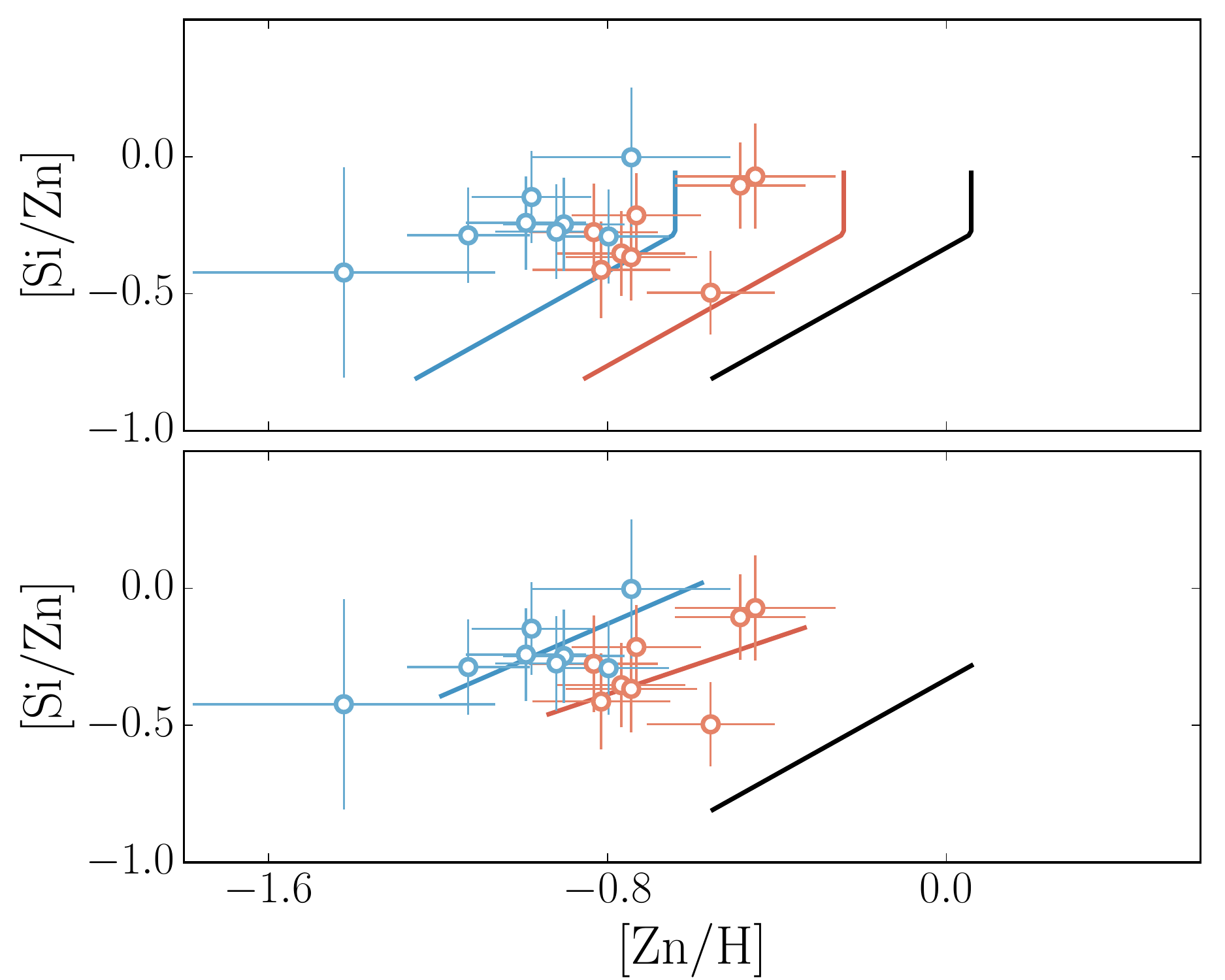}
  \caption{Blue and red data points are abundance measurements for
    the SMC and LMC from our COS sample. The black line is based on a
    fit to gas-phase abundances in the DNM of the MW
    \citep{Jenkins:2009ke}. 
    In the top panel, the red (LMC) and blue (SMC) lines are the MW line
    shifted to each galaxy's metallicity.
    In the bottom panel, red and blue lines are the LMC and SMC's linear depletion relations (see Sec.\ref{sec:depletiondist}). 
    Since all of the LMC and SMC points in the top panel are above their corresponding lines, gas-phase abundances in the MCs differ from those of the MW by more than just the relative abundance differences.}
  \label{fig:metalevo}
\end{figure}
Gas-phase abundance variations between pencil beam observations within a single galaxy can be succinctly summarized by the per-sightline depletion strength $F^*$ of the linear depletion relations of Section \ref{sec:depletiondist}.
Because these within-galaxy variations are quite large, with depletion-driven gas-phase abundance differences between sightlines of up to of order a decade, galaxy-to-galaxy comparisons should be not of individual sightlines but of galaxy-wide depletion relations. 
We will use the intrinsic depletion zero points and slopes of the linear depletion relations to summarize each galaxy's depletion relation.
While we do not have a quantitative "depletion-relation-relation" to describe variations between ensembles of pencil beam observations of different galaxies, we can use the changes we see from the MW to the LMC to the SMC as a qualitative depletion-relation sequence, at least to metallicities above that of the SMC.
Starting from the MW and moving towards lower metallicity, intrinsic depletion zero points decrease in magnitude, while depletion slopes remain approximately unchanged. 
The intrinsic depletion zero points of P and Zn, whose magnitude is already consistent with zero, do not change.

Comparing depletion relations between galaxies takes a (generally) prohibitive amount of information. 
Computing an abundance requires absorption-derived hydrogen columns, computing an intrinsic depletion requires galaxy-specific reference abundances (see Sec.\ref{sec:depletions}), and computing a depletion relation requires a representative sample of intrinsic depletions. 
In a more typical case, one instead has to work with a metallicity instead of a reference abundance for each element and less than ten sightlines per galaxy.
This was true of depletion studies of the MCs, remains true of depletion studies of other local galaxies \citep[e.g.][]{2013PASP..125.1455W,2014ApJ...795..109J}, and is, in a sense, intrinsically true of DLA and sub-DLA studies. 

A common exploratory technique in depletion studies of small samples involves making a comparison between a new measurement and some typical MW values for the abundance of a volatile element and the abundance ratio between the same volatile element and a refractory element. 
If one assumes that the MW depletion relation applies in the new measurement's galaxy, one can estimate each element's depletion and reference abundance. 
This type of exploratory analysis, if applied to the LMC or SMC, would be incorrect. 
We show an example of this type of exploratory analysis in the top panel of Figure \ref{fig:metalevo}.
The lines in this panel are the MW depletion relation shifted to the metallicity of the MW (black), LMC (red), and SMC (blue).
The diagonal portion of each line corresponds to a depletion strength ($F^*$) range of 0 (top right) to 1 (bottom left), and includes what \citet{1996ARA&A..34..279S} refer to as "warm disk" and "cool disk" depletion patterns. 
The vertical line extending upwards from $F^*=0$ corresponds to the "diffuse halo" depletion pattern range from the same work. 
These metallicity-shifted MW depletion relations are inconsistent with our observations.

In the bottom panel of Figure \ref{fig:metalevo}, the lines are our MC depletion relations from \ref{sec:depletiondist}. 
Examining these lines gives us a qualitative visual version of the earlier statement that intrinsic depletion zero points become smaller with decreasing metallicity. 
On a volatile abundance versus refractory-to-volatile ratio plot, a galaxy's intrinsic depletion zero point corresponds to the top right end of that galaxy's representative line.
Because, at least in these three galaxies, the intrinsic depletion zero point becomes less negative with decreasing metallicity, the top right end of each galaxy's line is shifted up and to the left as the metallicity decreases from solar (the MW) to 1/2 solar (the LMC) to 1/5 solar (the SMC).
The fraction of metals in dust decreases with decreasing metallicity.

\subsection{The Magellanic Clouds and damped Lyman-$\alpha$ systems}
\label{sec:evo:dla}
\begin{figure}
  \centering
  \includegraphics[width=\linewidth]{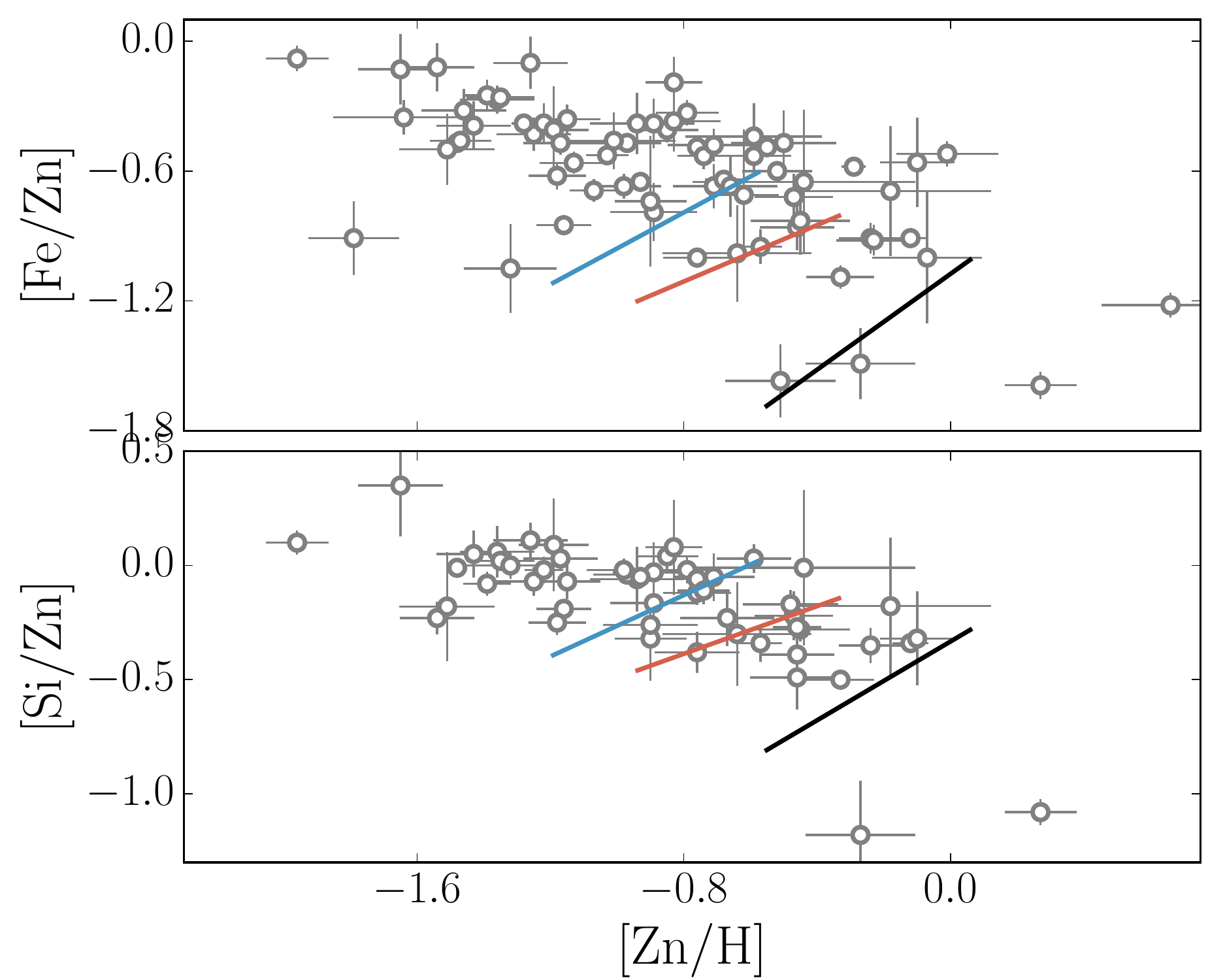}
  \caption{A comparison between abundances relative to solar in DLAs (data points,
    \citealt{Vladilo:2011hk},
    \citealt{2012ApJ...755...89R}) and the diffuse neutral media of the
    MW, LMC, and SMC (solid black, red, and blue lines, MW values from
    \citealt{Jenkins:2009ke}, LMC and SMC values from Section
    \ref{sec:depletiondist} of this work).}
  \label{fig:dla}
\end{figure}

DLAs are reservoirs of neutral gas with H I column densities greater
than $2\times 10^{20} \text{ cm}^{-2}$. For a review of DLA
properties, see \citet{2005ARA&A..43..861W}. Metal abundances in
DLAs range from approximately solar to less than $1/100$ of solar,
making the DLA population a valuable resource for understanding cosmic
chemical evolution. The conversion  between observed and intrinsic
abundances is a long standing problem in DLA studies, since one observes a single sightline and needs to deconvolve the effects of metallicity, abundance variations (especially $\alpha$-enrichment), and depletion \citep{1990ApJ...348...48P, 1995ApJ...445L..27S, 1997ApJ...484L...7K,2002ApJ...566...68P,2006MNRAS.370...43M,2012ApJ...755...89R,Vladilo:2011hk,2013A&A...560A..88D}.

The depletion part of this deconvolution problem is usually approached using some variation of the metallicity-scaled MW technique described in the previous section \citep[see, e.g.][]{2012ApJ...755...89R,2013A&A...560A..88D}.
Now that we have depletion relations from multiple galaxies over a range of metallicities, we are no longer limited to just the MW.
While we do not have a quantitative model for the evolution of galaxy-wide depletion relations, we do have the previous section's qualitative picture.
Figure \ref{fig:dla} compares the measured Fe-to-Zn and Si-to-Zn ratios and Zn abundances in DLAs with the corresponding ratios and abundances derived from the depletion relations of the MW, LMC and SMC. 
We first examine DLAs that have more positive Zn abundances or more negative refractory-to-volatile ratios than the SMC. 
The evolution of the Si-to-Zn ratio with the Zn abundance in our galaxy sample and in this DLA subsample appears to be the same.
The upper envelope of the Fe-to-Zn ratio is perhaps slightly less negative in this subsample of DLAs than in our galaxies, but the separation of the upper and lower envelopes along the direction of intra-galaxy dust evolution (i.e. parallel to the galaxy lines) is the same as the extent of the galaxy lines.
If we interpret a single DLA to be a pencil-beam-sized region in a larger system that can contain a variety of neutral medium conditions, the chemical evolution of the neutral ISM in these systems appears to be quite similar to that of the neutral ISM in the MCs or MW.

Interpreting the measurements with sub-SMC Zn abundances is more difficult. 
Applying a metallicity-scaled version of the MW depletion relation does not give us an adequate description of gas-phase abundances in the MCs, so metallicity-scaling the SMC depletion relation may not necessarily give us an adequate description of gas-phase abundances in low-metallicity DLAs. 
The shape of the upper and lower envelopes of the DLA abundances and abundance ratios in Figure \ref{fig:dla} broadly suggests that the concurrent metallicity and dust-to-metals ratio evolution that applies from the MW to the SMC continues to apply to even lower metallicities. 
By applying the previous paragraph's interpretation of a DLA as a single sightline through some system and assuming that the metallicity and dust-to-metals ratio of these systems evolves in a consistent way, one could come up with a more quantitative version of this statement.
We leave this exercise for the future.

\subsection{Gas-to-dust ratios in the Magellanic Clouds}

In Section \ref{sec:gdr}, we computed GDRs for the DNM of the
Magellanic Clouds from depletions. These GDRs are summarized in Figure
\ref{fig:gdr}. The DNM of each MC contains
(spatially separate) gas with a range of overall depletion
levels. In the figure, this level increases from left to right, as the
amount of dust per amount of gas increases. The GDRs of the least
depleted sightlines in the LMC and SMC DNM are 455-565 and 1540-2065; the
GDRs of the most depleted sightlines are 190-245 and 480-595. We have
compared these GDRs with values from the literature and found no
significant discrepancies.

\begin{table}
\caption{GDR measurements in the LMC and SMC}
\begin{tabular}{lcr}
  \hline
    {Location} & {GDR} & {Reference}\\
    \hline
    SMC DNM & 1600-2400 & \citet{2002ApJ...566..857T}\\ 
    SMC Wing & 480-720 & \citet{2003ApJ...594..279G}\\
    SMC Bar & 850-1275 & \citet{2003ApJ...594..279G}\\
    SMC DNM & 3000-4500 & {\citet{2004A&A...423..567B}}\\
    SMC, global average & 700 & \citet{2007ApJ...658.1027L}\\
    SMC Tail & 850-1550 & \citet{2009ApJ...690L..76G}\\
    SMC, CO-detected regions & 150-600 & \citet{2011ApJ...737...12L}\\
    SMC, global average & 1300-2180 & \citet{2014arXiv1406.6066G}\\
    SMC DNM & 480-2100 & This work\\
    \hline
    LMC DNM & 400-600 & \citet{2002ApJ...566..857T}\\
    LMC DNM & 200-300 & \citet{2003ApJ...594..279G}\\
    LMC supershells & 450-675 & \citet{2003ApJ...594..279G}\\
    LMC, global average & 270-405 & \citet{2008AJ....136..919B}\\
    LMC, CO-detected regions & 150-250 & \citet{2011ApJ...737...12L}\\
    LMC, global average & 300-400 & \citet{2014arXiv1406.6066G}\\
    LMC DNM & 190-565 & This work\\
    \hline
  \end{tabular}
  \tablecomments{Some of the sources
    give GDRs as factors of the MW GDR. These relative GDRs were
    converted to absolute values by assuming an MW GDR of 100-150.}
\label{tab:gdrs}
\end{table}

The literature values are given in Table \ref{tab:gdrs}. All of their uncertainties
have some overlap with the
our GDR ranges. This does not mean that our
GDR for any single depletion level is accurate. All of the values
in the table other than ours are averages over
different DNM conditions or, in some cases, multiple ISM phases. An average over a
(possibly) wide range of GDRs with unknown proportions of mass at each
GDR does not contain enough information to draw that sort of
conclusion. The decrease in GDR from averages over more diffuse
regions to averages over denser regions is heartening, though not conclusive.  

\section{Summary}
\label{sec:summary}
In order to study the gas and dust content of the LMC and SMC, we have obtained and analyzed 16 \emph{HST} COS NUV spectra and analyzed 85 archival \emph{FUSE} FUV spectra towards the Large and Small Magellanic Clouds.
From these spectra, we have measured P II, Zn II, Si II, Cr II, and Fe II column densities.
We have combined these measurements with H I and H$_2$ column densities and photospheric abundances of LMC and SMC stars from the literature to compute intrinsic elemental depletions, solid-phase elemental abundances, and GDRs.
We see large variations in gas-phase abundances, depletions, solid-phase abundances, and GDRs from sightline to sightline within each galaxy. 
The depletion variations can be accurately described using a linear depletion relation of the type defined for the MW by \citet{Jenkins:2009ke}, but with different linear coefficients. 

We find that the ISM properties of the LMC are, up to metallicity, quite similar to those of the MW while the ISM properties of the SMC have some significant differences. 
When we compare gas-phase abundances in the LMC and SMC with gas-phase abundances in damped Lyman-$\alpha$ systems, we find a considerable amount of overlap. 
Within this overlap region, the gas-phase abundance ranges of the MCs and DLAs are very similar.
Regardless of whether this similarity is meaningful, we caution against assuming that MW depletion patterns apply to low-metallicity systems. 

Below, we list a number of more specific results.
\begin{enumerate}
\item The minimum Si depletion in the SMC is consistent with zero while the Fe depletion along the same sightline is substantial, suggesting that the composition of some dust in the MW and SMC differs.
\item The volatile elements P and Zn have non-zero depletions along most of our sightlines through both MCs. The maximum P and Zn depletions in the MW, LMC, and SMC are the same to within the uncertainties. 
\item Si, Cr, and Fe are systematically less depleted in the SMC than in the LMC or MW. 
\item We find GDR ranges of 190-565 in the LMC and 480-2100 in the SMC. These ranges are broadly consistent with GDR values from the literature.
\item The GDRs of pencil-beam-sized parcels of neutral medium vary by a factor of 2 in the MW and LMC and by a factor of 5 in the SMC. 
The GDR variations within each galaxy suggest some combination of dust evolution in the neutral medium and rapid cycling of interstellar matter between the neutral medium and dense molecular clouds. 
\end{enumerate}

\begin{acknowledgements}
This research has made extensive use of the SIMBAD database, operated at CDS, Strasbourg, France; NASA's Astrophysics Data System Bibliographic Services; Astropy, a community-developed core Python package for Astronomy \citep{2013A&A...558A..33A}; NumPy and SciPy \cite{vanderWalt:dp}; IPython \citep{Perez:hy}; Matplotlib \citep{Hunter:ih}; and Cython \citep{Behnel:gs}.
We are grateful for financial support for this work from STScI grant HST-GO-13004.008 and NASA grant NNX13AE36G. 
Support for program GO-13004 was provided by NASA through grants from the Space Telescope Science Institute, which is operated by the Association of Universities for Research in Astronomy, Inc., under NASA contract NAS 5-26555.
\end{acknowledgements}

\appendix
\section{Analyzing a spectrum}
\label{ap:specfit}
The purpose of our model of a spectrum is to measure the total
column density $N^{\X_\ell}_{\text{MC}}$ of each species $\X_\ell$ in the
observation-appropriate Magellanic Cloud. We marginalize over, meaning integrate over the 
posterior probability distribution of, every other model parameter.
The other important high-level quantities are the (ISM absorption-free) continuum flux as a
function of wavelength $c(\lambda)$ and the distribution of $N^{\X_\ell}$ as a function of Doppler
velocity $N^{\X_\ell} (v)$. 
To convert these quantities to observables, we first
use the $N^{X_\ell}(v)$ to compute a transmittance function
$T(\lambda)$. 
Then, we convolve the product of the continuum and
transmittance with an instrumental line-spread function (LSF)
$L(\lambda, \lambda')$ to get a model flux $f(\lambda)$. 
This model flux is the mean of a multivariate normal (MVN) distribution,
from which we assume the data $y(\lambda)$ have been drawn. 
The covariance matrix $\Sigma$ of this MVN distribution is the sum of
a diagonal measurement uncertainty matrix $\Sigma_{\text{meas}}$ and a not necessarily diagonal
continuum uncertainty matrix $\Sigma_{\text{cont}}$. 

We fit for the ISM absorption-free continuum emission using Gaussian
process (GP) regression and prediction \citep{GPML}. 
The regression and prediction were done using the GP code \emph{george} \citep{1403.6015v1}.
Our main motivation for using a GP rather than, for
instance, splines is that the posterior predictive distribution of a
GP conditioned on observations can be used to exactly compute
the predicted values' covariant uncertainties. 
This covariant continuum uncertainty is an important part of our
model's error budget, particularly for high SNR measurements. 

We run the GP regression on a subset of the data that we find, by
visual inspection, to be free of ISM absorption, then compute the
predicted mean and covariance matrix of the full dataset. 
This subset comes as close to the ISM absorption as possible in order
to minimize the size of the prediction region. 
For the GP kernel, we use the sum of a square exponential kernel, a
quadratic kernel, and the measurement uncertainties of the ISM-free subset. 
We fix the lengthscale of the square exponential kernel to be 6 \AA\
and fit for the rest of the kernel parameters.

The dashed line in each panel of Figure~\ref{fig:fit:sk116} is an
example of a continuum fit. 
The uncertainty of the full spectrum model, which includes the continuum
uncertainty, is shown as a gray region in this figure. 

Instead of explicitly modeling the transmittance $T(\lambda)$, we
instead model the species' column density distributions
$N^{\X_\ell}(v)$ as a function of Doppler velocity and use a deterministic
non-linear transformation to convert that to $T(\lambda)$. 
The column density distributions are split into spatio-kinematic
groups $S_j$, which are velocity ranges that we \emph{a
  priori} assign to the MW, an I/HVC, or a MC, and the $S_j$ are
further split into components $t=1,\ldots, T_{S_j}$. 
Each component has a central velocity $\hat{v}^{S_j}_{t}$, a width
$b^{S_j}_t$, and, for each species $\X_\ell$, a fraction
$\alpha_{{S_j}_t}^{\X_\ell}$ of the group-level column density of $\X_\ell$.
The column density of species $\X_\ell$ in component $t$ is then
$\alpha_{{S_j}_t}^{\X_\ell} \times N^{\X_\ell}_{S_j}$; note that the
sum of the $\alpha$ for a single species over all of the components of
a group are equal to 1.

The column density distribution of a species is the sum of the group
level distributions, which are in turn the sum of the component level
distributions.
The column density distribution of a single component is
\begin{equation}
  \label{eq:single:profile}
   N_{{S_j}_t}^{\X_\ell} (v) = \frac{\sqrt{2}}{\sqrt{\pi} b^{S_j}_{t}} N^{\X_\ell}_{S_j}
    \alpha_{{S_j}_t}^{\X_\ell}
   \exp \left(-\left(\frac{v-\hat{v}^{S_j}_t}{b_t^{S_j}}\right)^2 \right)
\end{equation}
and the full column density distribution of a single element is
\begin{equation}
  \label{eq:profile:sum}
  N^{\X_\ell}(v) = \sum_{j=1}^J \, \sum_{t=1}^{T_j} N_{{S_j}_t}^{\X_\ell}(v).
\end{equation}

We use this two-level formalism because we believe that the
groups are physically meaningful while the components may or may not be.
The groups are physically meaningful because they correspond to
different galaxies with reasonably well-separated systemic velocities.
We believe that column density that has been assigned to a given group
most likely actually arises in the corresponding galaxy.
The components are not necessarily meaningful because,
firstly, our instrumental LSFs are broader than the expected
``individual'' components, secondly, components that are distinct in velocity are not necessarily distinct spatially, and thirdly, the fragmentation of a continuous density into discrete subcomponents is almost always not unique.
One consequence of the third point, particularly when the first point
applies, is that we cannot even define a unique \emph{number} of
components. 
This is especially true when our observations are a product of the
transmission function, which is a non-linear transformation and superposition
of the column density distributions, with the uncertain continuum. 
Instead of trying to choose a fixed number of unphysical components, we marginalize over it. 
This means that the dimension of our parameter space is not constant.

We generate samples from the posterior probability over this
variable-dimensional parameter space, using the
Reversible-Jump Markov Chain Monte Carlo 
(RJMCMC) algorithm \citep{Green:1995ut}.  
RJMCMC is a type of MCMC that can be applied to variable-dimensional parameter
spaces.
It works by considering each fixed-dimensional parameter space as a subspace of some larger, overarching parameter space.
One then defines different Markovian steps
for moving around within each fixed-dimensional subspace and between `adjacent' fixed-dimensional subspaces. 
Further details of this framework are beyond the scope of this appendix
and can be found in \citep{Green:1995ut}. 
We use these samples to build posterior probability distributions for each species'
column density in the LMC or SMC and the model flux at each
wavelength. The 16-th and 84-th percentiles of the column density posterior probability distributions 
are shown as error bars in Figure~\ref{fig:ioncolumns}. The 5-th, 16-th,
84-th, and 95-th percentiles of the model flux marginals
are shown as shaded regions in Figure~\ref{fig:fit:sk116}.
These marginals include measurement uncertainty, continuum
uncertainty, and the possibility of observationally similar but
physically different velocity component structures.

\section{Cloudy parameters}
\label{ap:cloudy}
Here, we provide the complete list of input parameters needed to
reproduce the photionization calculations described in section
\ref{sec:ioncorrections}. 
These calculations were done using version 13.02 of the Cloudy
photoionization code \citep{2013RMxAA..49..137F}. 
All of the models are of a constant-density cloud illuminated by a
star $10^{16}$ cm from the cloud surface. 
We compute models with cloud volume densities of 0.1, 1, 10, and 100
$\text{cm}^{-3}$. 
The cloud composition is given by the `abundances ism' command, scaled
to the appropriate metallicity (0.5 for the LMC, 0.2 for the SMC)
using the `metals grains' command. 
We run each cloud model for a B0-type and an O2-type illuminating
star of the appropriate metallicity using stellar atmosphere models from \citet{2003ApJS..146..417L}.

In addition to the illuminating star, the incident radiation field
includes the cosmic microwave background and the local MW radiation
field (`table ism') scaled by a factor of 4 in the LMC
\citep{2008AJ....136..919B} and 10 in the SMC
\citep{Sandstrom:2010ks}. We include the local cosmic ray background
using the `cosmic rays background' command.

\section{Elemental abundance meta-analysis}
\label{ap:meta}
In this appendix, we describe a multilevel model for the mean
elemental abundance $\mu$ of an element from a collection of possibly
biased studies with some repeated observations. 
A description of multilevel models and their application to meta-analysis problems can be found in Chapter 5 of \citet{Gelman:2013un}.
We can split the model into four blocks -- population level,
study level, individual star level, and study-star level. 

The population level includes the mean abundance $\mu$, the intrinsic
star-to-star variance $\sigma_{star}$, and the between-study variance
$\sigma_{study}$:
\begin{align}
  \mu &\sim \text{Normal}\left(\mu=\mu_\odot + Z_{MC}, \sigma^2=0.5^2\right)\\
  \sigma_{star} &\sim \text{Exponential}\left( \lambda=5 \right)\\
  \sigma_{study} &\sim \text{Exponential}\left( \lambda=5 \right),
\end{align}
where $\mu_\odot$ is the solar abundance of the relevant element,
$Z_{MC}$ is the metallicity of the relevant Magellanic Cloud, and the notation ``$x \sim \text{Distribution}$'', as in $\mu \sim \text{Normal}$, means that the specified distribution is our prior for $x$.
The study and star levels include the bias terms $\Delta_{study_j}$, for
study-wide effects, and $\Delta_{star_k}$, for star-to-star variations:
\begin{align}
  \Delta_{study_j} &\sim \text{Normal}\left(\mu=0, \sigma^2=\sigma_{study}^2 \right)\\
  \Delta_{star_k} &\sim \text{Normal}\left(\mu=0, \sigma^2=\sigma_{star}^2 \right)
\end{align}
The study-star level consists of observations $y_{j,k}$ and
observational precisions $\tau_{j,k}$:
\begin{align}
  \tau_{j,k} &\sim \text{Gamma}\left( \alpha=\alpha_{study_j},
    \beta=\beta_{study_j} \right)\\
  y_{j,k} &\sim \text{Normal}\left(
    \mu = \mu + \Delta_{study_j} + \Delta_{star_k},
    \sigma^2 = 1/\tau_{j,k}
    \right)\\
    \alpha_{study_j} &\equiv
    \left(\frac{\mu_{\tau_{study_j}}}{\sigma_{\tau_{study_j}}}\right)^2; \,
    \beta_{study_j} \equiv \frac{\mu_{\tau_{study_j}}} {\sigma_{\tau_{study_j}}^2},
\end{align}
where $\mu_{\tau_{study_j}}$ is the typical precision of the
measurements in study $j$ and $\sigma_{\tau_{study_j}}$ is the
dispersion in the typical precision.

We implemented this model using PyMC \citep{PyMC} and generated
samples from the posterior probability distribution using the No-U
Turn Sampler. We ran each element's MCMC chain for 500 burn-in steps and 5000
kept steps starting from the maximum a posteriori value, which, by
visual inspection, appeared to be a long-enough burn-in phase for
convergence. 
\bibliographystyle{apj}

\end{document}